\documentclass[letterpaper]{article} 
\usepackage[preprint]{aaai2027}
\usepackage[hyphens]{url}  
\usepackage{graphicx} 
\usepackage{natbib}  
\usepackage{caption} 
\usepackage{algorithm}
\usepackage{algorithmic}

\usepackage{newfloat}
\usepackage{listings}
\usepackage{booktabs}
\usepackage{multirow}
\usepackage{amsmath}
\usepackage{amsfonts}
\usepackage{tabularx} 
\usepackage{booktabs} 
\DeclareCaptionStyle{ruled}{labelfont=normalfont,labelsep=colon,strut=off} 
\floatstyle{ruled}
\newfloat{listing}{tb}{lst}{}
\floatname{listing}{Listing}

\title{SemBridge: Semantic Token Anchoring for Continuous-Latent\\
Autoregressive Speech Generation}
\author{
Hanke Xie\textsuperscript{\rm 1,2}\thanks{Work done during an internship at Soul AI Lab.}\equalcontrib,
Haopeng Lin\textsuperscript{\rm 2}\equalcontrib,
Jiale Qian\textsuperscript{\rm 2},
Dake Guo\textsuperscript{\rm 1},
Yuepeng Jiang\textsuperscript{\rm 1},
Zhichao Wang\textsuperscript{\rm 2},
Wenxiao Cao\textsuperscript{\rm 2},
Jingbin Hu\textsuperscript{\rm 1},
Guobin Ma\textsuperscript{\rm 1},
Wenhao Li\textsuperscript{\rm 1},
Huakang Chen\textsuperscript{\rm 1},
Chengyou Wang\textsuperscript{\rm 1},
Ming Tao\textsuperscript{\rm 2},
Zhonghua Fu\textsuperscript{\rm 1},
Lei Xie\textsuperscript{\rm 1}\corresponding,
Xinsheng Wang\textsuperscript{\rm 2}\corresponding
}

\affiliations{
\textsuperscript{\rm 1}Audio, Speech and Language Processing Group 
(ASLP@NPU), School of Software \\ Northwestern Polytechnical University, Xi'an, China\\
\textsuperscript{\rm 2}Soul AI Lab, China\\
hkxie@mail.nwpu.edu.cn, lxie@nwpu.edu.cn,
wangxinsheng@soulapp.cn
}

\newcommand{\methodname}{SemBridge}
\newcommand{\tbd}{\textcolor{gray}{TBD}}

\begin{document}

\maketitle


\begin{abstract}
Continuous-latent autoregressive speech generation has emerged as a promising
alternative to discrete-token modeling by avoiding quantization loss and
preserving richer acoustic information. However, continuous acoustic targets
do not expose linguistic structure as explicit token-level prediction targets. Consequently, the autoregressive language model (LM) must acquire linguistic structure indirectly through acoustic prediction, which can compromise the content fidelity of generated speech.
We propose \textbf{SemBridge}, a training-only semantic-token anchoring
framework for continuous-latent autoregressive speech generation. SemBridge
uses discrete semantic tokens to directly supervise autoregressive LM states
and employs a Semantic-Aligned Acoustic VAE to organize the continuous target
space under the same semantic reference. The semantic supervision is used
only during training, while inference remains entirely continuous.
We evaluate SemBridge on zero-shot text-to-speech (TTS) and score-conditioned
singing voice synthesis (SVS). Across multiple benchmarks, SemBridge improves content accuracy, as measured
by word and character error rates (WER/CER), while maintaining competitive
speaker similarity and perceptual quality.
Experimental results demonstrate that explicit semantic-token supervision
for autoregressive state learning is an effective and general direction for continuous speech generation. 
\begin{links}
    \link{Code}{https://github.com/ASLP-lab/SemBridge}
    \link{Demo}{https://tiamojames.github.io/SemBridge_Demo}
\end{links}

\end{abstract}

\section{Introduction}

Autoregressive (AR) models have emerged as a dominant paradigm for speech synthesis.~\cite{wang2023valle,anastassiou2024seedtts}. Most AR speech generation systems adopt a language-modeling formulation,
representing speech as discrete token sequences and formulating generation as standard
next-token prediction. Recent AR systems further advance discrete speech representations by introducing low-rate and semantically structured speech tokens through pretrained speech representations or hierarchical semantic--acoustic modeling~\cite{cosyvoice2,du2025cosyvoice3,hu2026qwen3,zhang2023speechtokenizer}.
These designs provide compact and linguistically meaningful prediction targets, reducing the text--speech semantic gap and improving autoregressive modeling. However, discrete representations inevitably introduce quantization constraints, limiting their capacity to preserve fine-grained acoustic details required for high-fidelity speech generation.
Continuous representations avoid quantization loss while providing a richer and more
expressive acoustic target space. Consequently, continuous-latent
autoregressive generation has emerged as a promising direction for high-fidelity speech synthesis~\cite{meng2025melle,jia2025ditar,xia2026kalle}.

Continuous-latent autoregressive models generate speech by predicting
continuous acoustic representations, such as mel-spectrograms or learned
acoustic latents, often using diffusion- or flow-based next-patch generation
to model complex acoustic distributions
~\cite{jia2025ditar,peng2026vibevoice,zhou2025voxcpm}. These continuous
targets preserve rich acoustic information while implicitly encoding linguistic content.
Unlike discrete semantic tokens, however, they do not provide explicit
semantic targets at each generation step. Consequently, the
autoregressive LM is required to infer linguistic structure indirectly from the continuous generation objective, without explicit token-level semantic supervision.

Recent studies have explored various strategies for incorporating semantic
information into continuous speech generation. Existing approaches mainly
leverage continuous semantic representations extracted from pretrained speech
models to enhance either target representations or autoregressive hidden
states~\cite{wang2026semavoice,an2025melatts}. While these methods demonstrate the
importance of semantic modeling for continuous generation, they formulate semantic supervision as continuous feature alignment. In contrast, discrete semantic tokens organize linguistic content into discrete semantic units and provide explicit prediction targets for autoregressive language modeling. Such discretization makes clusterable linguistic structure explicit while suppressing linguistically irrelevant variation retained in continuous speech features ~\cite{hsu2021hubert,nguyen2022discrete}. 
However, leveraging discrete semantic tokens to explicitly supervise continuous-latent autoregressive states remains underexplored.
Other approaches introduce hierarchical or semi-discrete architectures to combine semantic and acoustic modeling~\cite{zhou2025voxcpm,DBLP:journals/corr/abs-2512-19090};
however, they modify the original continuous generation interface by
introducing additional semantic generation components.

To address this gap, we introduce \methodname{}, a training-only
semantic-token anchoring framework for continuous-latent autoregressive speech
generation. At its core, \methodname{} uses discrete semantic-token labels to
provide direct semantic supervision for autoregressive LM states, while
continuous latent patches remain the sole generated and recurrent variables.
A complementary Semantic-Aligned Acoustic VAE (SA-VAE) further aligns continuous
acoustic latents with token-level embeddings from the same semantic tokenizer,
establishing a shared semantic reference for latent representation learning and
state supervision.

We evaluate \methodname{} primarily on zero-shot text-to-speech (TTS) and
further assess its transferability to score-conditioned singing voice
synthesis (SVS)~\cite{liu2021diffsinger,zhang2021visinger,
qian2026soulxsinger}. Across multiple TTS and SVS benchmarks, \methodname{}
consistently reduces WER/CER while maintaining competitive speaker similarity
and perceptual quality. Controlled ablations show that semantic-token
anchoring provides complementary gains beyond target-space alignment alone,
while layer-wise analyses reveal a trade-off between content
accuracy and synthesis quality.

Our main contributions are summarized as follows:
\begin{itemize}
    \item We propose \methodname{}, a training-only semantic-token anchoring framework that directly supervises autoregressive LM states with discrete semantic-token labels while preserving continuous-only generation during inference.

    \item We introduce a unified semantic supervision scheme based on a shared frozen semantic tokenizer: token-level embeddings align continuous acoustic latents through SA-VAE, while discrete semantic-token labels explicitly supervise autoregressive LM hidden states.

    \item Experiments on zero-shot TTS and score-conditioned SVS show that \methodname{} consistently improves WER/CER while maintaining competitive speaker similarity and perceptual quality. Ablations further confirm the complementary benefits of latent alignment and state anchoring and reveal a trade-off effect of anchoring depth.
\end{itemize}




\section{Related Work}

\paragraph{Discrete-Token Autoregressive Speech Generation.}
Modern LM-based speech systems commonly represent speech as discrete token
sequences produced by semantic or neural-codec tokenizers. AudioLM, VALL-E,
and Seed-TTS formulate speech generation as autoregressive sequence modeling
over discrete speech representations, enabling scalable generation and
zero-shot adaptation~\cite{borsos2022audiolm,wang2023valle,
anastassiou2024seedtts}. Recent work increasingly explores low-rate and
semantically informed tokens to shorten speech sequences and strengthen
linguistic modeling. SpeechTokenizer distributes semantic and acoustic
information across multiple quantization levels, while WavTokenizer explores
a compact single-codebook representation
~\cite{zhang2023speechtokenizer,ji2024wavtokenizer}. Self-supervised speech
studies further show that learned representations exhibit clusterable
phonetic structure, while discrete-unit analyses demonstrate that
discretization can suppress linguistically irrelevant variation and provide
effective prediction targets for spoken language modeling
~\cite{hsu2021hubert,nguyen2022discrete}. Nevertheless, quantization
introduces an information bottleneck that can limit the preservation of
fine-grained acoustic details.

\paragraph{Continuous-Latent Autoregressive Speech Generation.}
Continuous-latent approaches avoid discrete quantization by directly modeling
mel-spectrograms, learned acoustic latents, or latent distributions. MELLE
autoregressively generates continuous mel-spectrogram frames, while KALL-E
models the distribution of future continuous latents
~\cite{meng2025melle,xia2026kalle}. Recent methods further combine causal
sequence modeling with diffusion- or flow-based local generation. DiTAR
separates inter-patch dependency modeling from intra-patch acoustic generation
through a causal LM and a local diffusion transformer
~\cite{jia2025ditar}, while VibeVoice uses low-rate continuous
representations and next-patch diffusion for long-form multi-speaker
generation~\cite{peng2026vibevoice}. These approaches preserve an expressive
continuous acoustic target space, but their autoregressive states are trained
primarily through continuous generation objectives.

\paragraph{Semantic Guidance for Continuous Autoregressive Generation.}
Recent methods introduce semantic guidance into continuous generation through
pretrained speech representations. SemaVoice aligns continuous VAE latents
with high-level features from a frozen speech foundation model, improving
semantic structure in the acoustic target space before autoregressive
training~\cite{wang2026semavoice}. MELA-TTS instead regularizes
autoregressive states by aligning them with continuous ASR features
~\cite{an2025melatts}. These methods transfer semantic information through
continuous feature alignment at the target and predictor levels,
respectively. VoxCPM adopts a semi-discrete semantic-prosodic bottleneck
together with a residual pathway, retaining an explicit semantic
modeling branch during inference~\cite{zhou2025voxcpm}. 


\begin{figure*}[t]
\centering
\includegraphics[width=1.\textwidth]{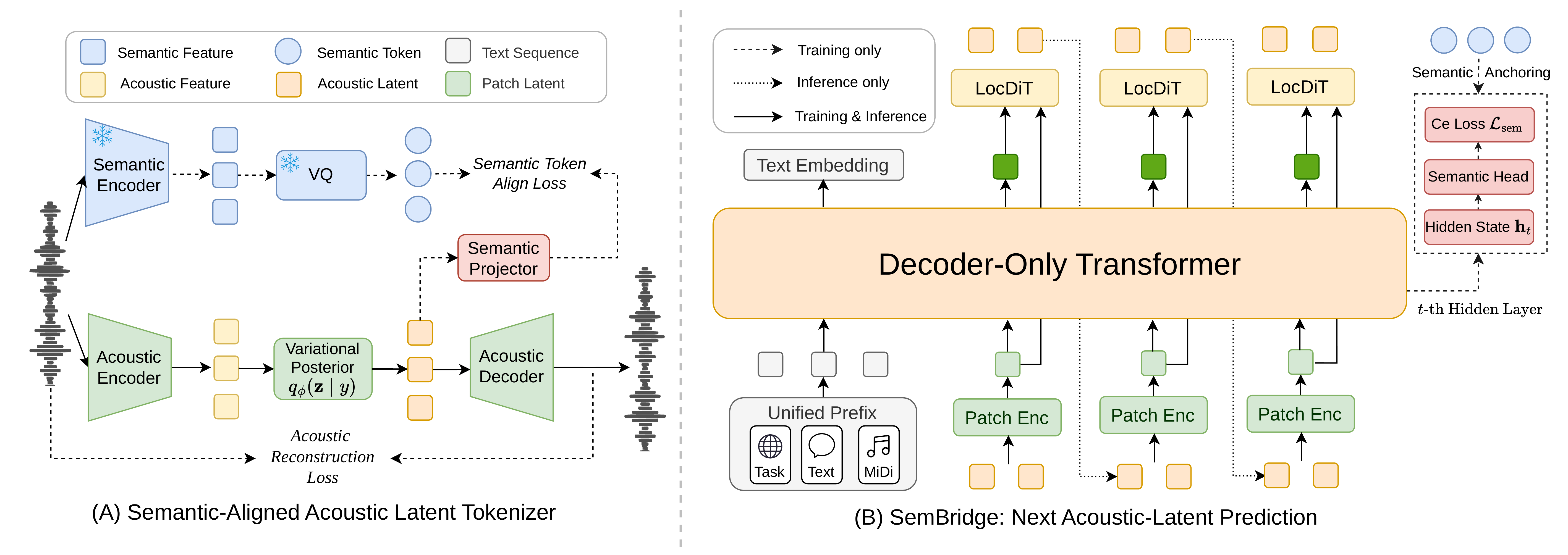}
\caption{
Overview of the two-stage \methodname{} framework.
(A) Stage I trains SA-VAE to reconstruct waveforms while aligning continuous
acoustic patches with embeddings from a frozen semantic tokenizer; their token
IDs are retained as Stage-II targets.
(B) Stage II trains a continuous-latent autoregressive generator: a causal LM
encodes symbolic conditions and preceding acoustic patches, LocDiT predicts the
next patch, and a semantic head supervises a LM hidden layer under the same
causal shift.
}
\label{fig:sembridge_two_stage}
\end{figure*}

\section{Method}
\label{sec:method}

\subsection{Overview}
\label{sec:overview}

\methodname{} is built on a continuous-latent autoregressive backbone that
generates acoustic latent patches conditioned on symbolic inputs and preceding
acoustic history. Its core mechanism is semantic-token anchoring, which uses
discrete semantic-token labels to provide direct supervision for selected
autoregressive LM states. A complementary Semantic-Aligned Acoustic VAE
(SA-VAE) organizes the continuous acoustic target space under the same
semantic reference. Together, the two objectives introduce semantic structure
into both the predictive LM states and the continuous targets they model,
while retaining continuous latent patches as the sole generated and recurrent
variables.

We use an open-source pretrained GLM-4 tokenizer\cite{zeng2024glm4voice} operating at
\(12.5\)~Hz. Given a training waveform \(\mathbf{x}\), the tokenizer produces
a sequence of token-level semantic embeddings and their corresponding
discrete token IDs:
\begin{equation}
\left(
\mathbf{E},
\mathbf{S}
\right)
=
\mathcal{S}_{\mathrm{GLM}}(\mathbf{x}),
\mathbf{E}
=
\{\mathbf{e}_t\}_{t=1}^{T},
\mathbf{S}
=
\{s_t\}_{t=1}^{T},
\label{eq:shared_semantic_interface}
\end{equation}
where \(\mathbf{e}_t\) denotes the continuous embedding associated with the
semantic token \(s_t\), and
\(s_t\in\{1,\ldots,V\}\) with vocabulary size \(V=16{,}384\).
The embeddings and token IDs provide two complementary forms of supervision:
\(\mathbf{e}_t\) is used for continuous target-space alignment, whereas
\(s_t\) serves as the categorical target for LM-state anchoring.

The SA-VAE encoder produces frame-level continuous acoustic latents
\(\mathbf{A}=\{\mathbf{a}_n\}_{n=1}^{2T}\) at \(25\)~Hz. We group every two
consecutive frames into one autoregressive acoustic patch:
\begin{equation}
\mathbf{z}_t
=
\operatorname{Concat}
\left(
\mathbf{a}_{2t-1},
\mathbf{a}_{2t}
\right),
\qquad
t=1,\ldots,T.
\label{eq:acoustic_patch_construction}
\end{equation}
The resulting acoustic patch sequence
\(\mathbf{Z}=\{\mathbf{z}_t\}_{t=1}^{T}\) therefore has the same \(12.5\)-Hz
rate as the semantic tokenizer outputs. This establishes a one-to-one temporal
correspondence:
\begin{equation}
\mathbf{z}_t
\longleftrightarrow
\left(
\mathbf{e}_t,
s_t
\right).
\label{eq:patch_semantic_correspondence}
\end{equation}
Accordingly, each continuous patch is aligned with a token-level embedding
during SA-VAE training, and the LM state responsible for predicting that
patch is supervised by the corresponding discrete token ID during generator
training.

Training proceeds in two stages as shown in Figure \ref{fig:sembridge_two_stage}. Stage~I trains SA-VAE with acoustic
reconstruction and semantic target-alignment objectives. Stage~II freezes
SA-VAE and the GLM-4 semantic tokenizer, and trains the continuous
autoregressive generator with its original generation objectives together
with semantic-token anchoring. Although SA-VAE is trained first, the defining
component of \methodname{} is the Stage-II token-level supervision applied
directly to autoregressive LM states.

\subsection{Continuous-Latent Autoregressive Backbone}
\label{sec:continuous_ar}

Given the continuous acoustic patch sequence
\(\mathbf{Z}=\{\mathbf{z}_1,\ldots,\mathbf{z}_T\}\) produced by the frozen
SA-VAE, SemBridge autoregressively models speech generation conditioned on a
symbolic input \(\mathbf{c}\) and an optional acoustic prompt
\(\mathbf{r}\):
\begin{equation}
p(\mathbf{Z}\mid\mathbf{c},\mathbf{r})
=
\prod_{t=1}^{T}
p
\left(
\mathbf{z}_t
\mid
\mathbf{c},
\mathbf{r},
\mathbf{z}_{<t}
\right).
\label{eq:ar_factorization}
\end{equation}

\paragraph{Autoregressive Context Modeling.}

Following recent continuous-latent autoregressive frameworks
~\cite{jia2025ditar}, the generator consists of PatchEnc, a causal LM, and a
local diffusion transformer, denoted as LocDiT. PatchEnc maps the acoustic
prompt and preceding continuous patches into acoustic context representations,
which are combined with the symbolic conditions and processed by the causal
LM. At the position responsible for predicting \(\mathbf{z}_t\), the hidden
state produced by Transformer layer \(\ell\) is:
\begin{equation}
\mathbf{h}_t^{(\ell)}
=
\mathcal{H}_{\theta}^{(\ell)}
\left(
\mathbf{c},
\mathbf{r},
\mathbf{z}_{<t}
\right),
\qquad
\ell=1,\ldots,L.
\label{eq:causal_hidden_state}
\end{equation}
The causal construction ensures that \(\mathbf{h}_t^{(\ell)}\) only depends on
the symbolic conditions, prompt context, and preceding acoustic patches. The
final-layer state provides the global condition for continuous next-patch
generation, while a selected hidden layer is additionally supervised by the
semantic-token anchoring objective described in
Section~\ref{sec:anchoring}Semantic-Token Anchoring. 

\paragraph{Conditional Continuous Patch Generation.}

The normalized final LM state
\begin{equation}
\mathbf{g}_t
=
\operatorname{RMSNorm}
\left(
\mathbf{h}_t^{(L)}
\right)
\label{eq:global_condition}
\end{equation}
conditions LocDiT together with the preceding acoustic patch
\(\mathbf{z}_{t-1}\). Rather than directly regressing
\(\mathbf{z}_t\), LocDiT models its conditional distribution using flow
matching:
\begin{equation}
\mathcal{L}_{\mathrm{FM}}
=
\mathbb{E}
\left[
\left\|
v_{\Theta}
\left(
\widetilde{\mathbf{z}}_{t,\tau},
\tau;
\mathbf{g}_t,
\mathbf{z}_{t-1}
\right)
-
\mathbf{v}^{\star}_t
\right\|_2^2
\right],
\label{eq:flow_matching}
\end{equation}
where \(\tau\) denotes the flow time,
\(\widetilde{\mathbf{z}}_{t,\tau}\) is the corresponding intermediate flow
state, and \(\mathbf{v}^{\star}_t\) is the target velocity field. A lightweight
stop predictor attached to the final LM state determines sequence
termination.


\subsection{Semantic-Token Anchoring}
\label{sec:anchoring}

\paragraph{Causally Aligned Semantic Prediction.}

Let
\(\mathbf{h}_t^{(\ell)}\) denote the post-block hidden state produced by
Transformer layer \(\ell\) at the autoregressive position responsible for
predicting the next continuous patch \(\mathbf{z}_t\):
\begin{equation}
\mathbf{h}_t^{(\ell)}
=
\mathcal{H}_{\theta}^{(\ell)}
\left(
\mathbf{c},
\mathbf{r},
\mathbf{z}_{<t}
\right),
\qquad
\ell=1,\ldots,L.
\label{eq:causal_hidden_state2}
\end{equation}
Because the GLM-4 semantic tokenizer and the autoregressive acoustic patches
share the same \(12.5\)-Hz temporal rate, the semantic token \(s_t\) is
associated with the same target patch \(\mathbf{z}_t\). Under the causal shift
used for next-patch prediction, the state \(\mathbf{h}_t^{(\ell)}\) has access
only to the symbolic conditions, prompt context, and preceding acoustic
patches \(\mathbf{z}_{<t}\). It does not observe either the target patch
\(\mathbf{z}_t\) or its semantic label \(s_t\).

The same autoregressive state is therefore trained to support two
complementary predictions:
\begin{equation}
\mathbf{h}_t^{(L)}
\longrightarrow
\mathbf{z}_t,
\qquad
\mathbf{h}_t^{(\ell^\star)}
\longrightarrow
s_t,
\label{eq:dual_prediction_targets}
\end{equation}
where the final-layer state conditions continuous next-patch generation,
while a selected intermediate or final layer
\(\ell^\star\) is anchored by the corresponding semantic-token label.
Semantic tokens are used only as supervision targets and never enter the
autoregressive context.

\paragraph{Depth-Selective Token Classification.}

We attach a lightweight semantic prediction head
\(f_{\phi}\) to the post-block hidden state at a selected Transformer depth
\(\ell^\star\). The conditional semantic-token distribution is:
\begin{equation}
q_{\phi}
\left(
s_t
\mid
\mathbf{h}_t^{(\ell^\star)}
\right)
=
\operatorname{softmax}
\left(
f_{\phi}
\left(
\mathbf{h}_t^{(\ell^\star)}
\right)
\right).
\label{eq:semantic_prediction}
\end{equation}
Unlike continuous feature alignment, this objective does not require the
autoregressive state to reproduce the full geometry of a teacher
representation. Instead, it assigns an explicit semantic classification
target to each state responsible for continuous next-patch prediction.

The anchoring objective is computed over all valid acoustic prediction
positions:
\begin{equation}
\mathcal{L}_{\mathrm{sem}}
=
-
\frac{1}{|\mathcal{T}|}
\sum_{t\in\mathcal{T}}
\log
q_{\phi}
\left(
s_t
\mid
\mathbf{h}_t^{(\ell^\star)}
\right),
\label{eq:semantic_loss}
\end{equation}
where \(\mathcal{T}\) denotes the set of valid autoregressive acoustic
positions after excluding padded positions. We allow the attachment depth
\(\ell^\star\) to vary and study its effect experimentally.

\paragraph{Joint Continuous and Semantic Objectives.}

The semantic-token objective is jointly optimized with the original
continuous-latent generation losses:
\begin{equation}
\mathcal{L}_{\mathrm{gen}}
=
\mathcal{L}_{\mathrm{FM}}
+
\lambda_{\mathrm{sem}}
\mathcal{L}_{\mathrm{sem}}
+
\lambda_{\mathrm{stop}}
\mathcal{L}_{\mathrm{stop}}.
\label{eq:generator_objective}
\end{equation}
The flow-matching objective trains continuous next-patch generation, while
\(\mathcal{L}_{\mathrm{sem}}\) updates the semantic head and the
autoregressive LM states. The SA-VAE and GLM-4 semantic tokenizer remain
frozen throughout Stage~II. Consequently, the semantic objective acts as a
state-level training constraint without changing the acoustic target or
introducing discrete variables into the recurrent generation history.

\paragraph{Continuous-Only Inference.}

At inference, neither semantic-token labels nor the semantic prediction head
are required. Generation follows the original continuous
PatchEnc--LM--LocDiT pathway:
\begin{equation}
p(\mathbf{Z}\mid\mathbf{c},\mathbf{r})
=
\prod_{t=1}^{T}
p
\left(
\mathbf{z}_t
\mid
\mathbf{c},
\mathbf{r},
\mathbf{z}_{<t}
\right).
\label{eq:continuous_inference}
\end{equation}
Continuous latent patches remain the only generated variables and recurrent
acoustic history. The semantic-token branch therefore changes how the
autoregressive states are learned, rather than how speech is generated.

\subsection{Target-Space Alignment with SA-VAE}
\label{sec:savae}
Complementary to state-level semantic-token anchoring, we introduce a
Semantic-Aligned Acoustic VAE (SA-VAE) to improve the semantic correspondence
of the continuous acoustic target space. SA-VAE remains a continuous Gaussian
VAE and uses a frozen semantic tokenizer as its semantic reference. Given an
input waveform \(\mathbf{x}\), the tokenizer extracts token-level semantic
embeddings
\(\mathbf{e}=\mathcal{S}_{\mathrm{feat}}(\mathbf{x})\). A lightweight projection maps the continuous acoustic latent representations
into the semantic feature space, and semantic correspondence is optimized through:
\begin{equation}
\mathcal{L}_{\mathrm{align}}
=
\mathcal{D}_{\mathrm{sem}}
\left(
\mathcal{P}_{z}(\mathbf{z}),
\operatorname{sg}
\left[
\mathbf{e}
\right]
\right),
\label{eq:semantic_alignment_loss}
\end{equation}
where \(\mathcal{P}_{z}\)denote lightweight alignment
mappings, \(\operatorname{sg}(\cdot)\) stops gradients through the frozen
semantic tokenizer, and \(\mathcal{D}_{\mathrm{sem}}\) combines cosine
alignment with a weak \(L_1\) constraint.

The complete Stage-I objective is:
\begin{equation}
\mathcal{L}_{\mathrm{SA\text{-}VAE}}
=
\mathcal{L}_{\mathrm{VAE}}
+
\lambda_{\mathrm{align}}
\mathcal{L}_{\mathrm{align}} .
\label{eq:savae_objective}
\end{equation}
This objective encourages patch-level semantic correspondence without imposing
a discrete bottleneck on the acoustic representation. After Stage~I training,
SA-VAE is frozen and provides continuous acoustic targets for generator
training.

Target-space alignment and state-level anchoring use different forms of
supervision from the same semantic tokenizer. SA-VAE aligns
\(\mathbf{z}_t\) with the continuous embedding \(\mathbf{e}_t\), whereas
semantic-token anchoring uses the corresponding discrete ID \(s_t\) to
supervise the LM state responsible for predicting \(\mathbf{z}_t\). Their
shared semantic identity and temporal correspondence make the two objectives
complementary.

\subsection{Task Conditioning for TTS and SVS}
\label{sec:task_conditioning}

Semantic-token anchoring is applied to the autoregressive LM states responsible
for continuous next-patch prediction and is shared across TTS and SVS. TTS
uses text and language conditions, whereas score-conditioned SVS additionally
incorporates lyrics and score-derived pitch and duration; both tasks otherwise
share the same continuous acoustic representation, PatchEnc--LM--LocDiT
backbone, and anchoring objective. SVS therefore provides a more demanding
extension for examining whether semantic-token supervision improves lyric
intelligibility under explicit musical constraints without sacrificing the
fine-grained acoustic variation and expressiveness of singing.

\begin{table*}[t]
\centering
\setlength{\tabcolsep}{1.5pt}
\renewcommand{\arraystretch}{0.95}

\newcommand{\ccsim}[2]{\makebox[2.8em][r]{#1}\,\textbar\,\makebox[2.8em][l]{#2}}

\resizebox{\linewidth}{!}{%
\begin{tabular}{@{}lcc|ccc|cccc@{}}
\toprule
\multirow{3}{*}{\textbf{Model}} & \multirow{3}{*}{\textbf{Params}} & \multirow{3}{*}{\textbf{Data (h)}} & \multicolumn{7}{c}{\textbf{Content Consistency} $\downarrow$ \;|\; \textbf{Speaker Similarity} $\uparrow$} \\
\cmidrule(lr){4-10}
& & & \multicolumn{3}{c|}{\textbf{Seed-TTS-Eval}} & \multicolumn{4}{c}{\textbf{CV3-EVAL}} \\
& & & \textbf{ZH} & \textbf{EN} & \textbf{ZH-Hard} & \textbf{ZH} & \textbf{EN} & \textbf{ZH-Hard} & \textbf{EN-Hard} \\
\midrule

\multicolumn{10}{l}{\textit{Reported and open-source systems}} \\
F5-TTS & 0.3B & 100k & \ccsim{1.53}{0.760} & \ccsim{2.00}{0.670} & \ccsim{8.67}{0.713} & \ccsim{5.47}{--} & \ccsim{8.90}{--} & \ccsim{--}{--} & \ccsim{--}{--} \\
CosyVoice2 & 0.5B & 170k & \ccsim{1.38}{0.757} & \ccsim{3.09}{0.659} & \ccsim{\underline{6.83}}{0.724} & \ccsim{4.08}{--} & \ccsim{6.32}{--} & \ccsim{12.58}{0.726} & \ccsim{11.96}{0.667} \\
CosyVoice3-1.5B & 1.5B & 1.0M & \ccsim{1.12}{\underline{0.781}} & \ccsim{2.22}{0.720} & \ccsim{\textbf{5.83}}{\textbf{0.758}} & \ccsim{3.91}{--} & \ccsim{4.99}{--} & \ccsim{\underline{9.77}}{\underline{0.785}} & \ccsim{10.55}{\textbf{0.761}} \\
IndexTTS2 & 1.5B & -- & \ccsim{1.03}{0.765} & \ccsim{2.23}{0.706} & \ccsim{7.12}{\underline{0.755}} & \ccsim{--}{--} & \ccsim{--}{--} & \ccsim{--}{--} & \ccsim{--}{--} \\
HiggsAudio-v2 & 3B & 10M & \ccsim{1.50}{0.740} & \ccsim{2.44}{0.677} & \ccsim{55.07}{0.656} & \ccsim{3.89}{--} & \ccsim{5.24}{--} & \ccsim{14.14}{\textbf{0.786}} & \ccsim{9.04}{\underline{0.759}} \\

\midrule

\multicolumn{10}{l}{\textit{Continuous-latent autoregressive systems}} \\
DiTAR & 0.6B & -- & \ccsim{1.02}{0.753} & \ccsim{\textbf{1.69}}{\underline{0.735}} & \ccsim{--}{--} & \ccsim{--}{--} & \ccsim{--}{--} & \ccsim{--}{--} & \ccsim{--}{--} \\
MELA-TTS & 0.8B & 170k & \ccsim{\underline{0.95}}{0.720} & \ccsim{2.45}{0.590} & \ccsim{7.75}{0.710} & \ccsim{--}{--} & \ccsim{--}{--} & \ccsim{--}{--} & \ccsim{--}{--} \\
VoxCPM-Emilia & 0.6B & 100k & \ccsim{1.11}{0.740} & \ccsim{2.34}{0.681} & \ccsim{12.46}{0.698} & \ccsim{4.47}{--} & \ccsim{5.23}{--} & \ccsim{22.20}{0.626} & \ccsim{10.00}{0.626} \\
VoxCPM & 0.6B & 1.8M & \ccsim{\textbf{0.93}}{0.772} & \ccsim{1.85}{0.729} & \ccsim{8.87}{0.730} & \ccsim{\underline{3.40}}{--} & \ccsim{\textbf{4.04}}{--} & \ccsim{12.90}{0.661} & \ccsim{\underline{7.89}}{0.643} \\
SemaVoice & 1.5B & 100k & \ccsim{1.32}{0.728} & \ccsim{1.91}{0.657} & \ccsim{9.37}{0.687} & \ccsim{--}{--} & \ccsim{--}{--} & \ccsim{--}{--} & \ccsim{--}{--} \\
VibeVoice$^{*}$ & 1.5B & -- & \ccsim{2.78}{0.686} & \ccsim{5.28}{0.588} & \ccsim{29.43}{0.583} & \ccsim{5.16}{\underline{0.702}} & \ccsim{6.96}{\underline{0.560}} & \ccsim{15.94}{0.637} & \ccsim{\underline{7.89}}{0.586} \\
VoxCPM2 & 2B & 2M & \ccsim{0.97}{\textbf{0.795}} & \ccsim{1.84}{\textbf{0.753}} & \ccsim{8.13}{0.753} & \ccsim{3.65}{--} & \ccsim{5.00}{--} & \ccsim{\textbf{8.55}}{--} & \ccsim{8.48}{--} \\

\midrule



\textbf{SemBridge} & 0.8B & 120k & \ccsim{\underline{0.95}}{0.758} & \ccsim{\underline{1.81}}{0.699} & \ccsim{9.79}{0.717} & \ccsim{\textbf{3.34}}{\textbf{0.757}} & \ccsim{\underline{4.22}}{\textbf{0.658}} & \ccsim{10.58}{0.717} & \ccsim{\textbf{6.35}}{0.619} \\

\bottomrule
\end{tabular}%
}
\caption{
Objective evaluation on Seed-TTS-Eval and CV3-EVAL.
Each result is reported as content consistency error~$\mid$~speaker
similarity, where Chinese subsets use CER, English subsets use WER, and
speaker similarity is measured by SIM.
Lower content error and higher SIM indicate better performance.
``--'' denotes unavailable results.
Bold and underlined values indicate the best and second-best results,
respectively. Tied second-best results are both underlined.
$^{*}$ denotes results obtained through our evaluation.
}
\label{tab:speech_generation_main}
\end{table*}

\begin{table}[t]
\centering
{\footnotesize
\setlength{\tabcolsep}{1.0pt}
\renewcommand{\arraystretch}{0.95}
\resizebox{\columnwidth}{!}{%
\begin{tabular}{@{}l|cc|cc|ccc@{}}
\toprule
\textbf{Model} & \multicolumn{2}{c|}{\textbf{ZH}} & \multicolumn{2}{c|}{\textbf{EN}} & \multicolumn{3}{c}{\textbf{ZH-Hard}} \\
& CER $\downarrow$ & SIM $\uparrow$ & WER $\downarrow$ & SIM $\uparrow$ & CER $\downarrow$ & SIM $\uparrow$ & UTMOS $\uparrow$ \\
\midrule

\multicolumn{8}{@{}l}{\textit{Continuous-AR Baselines}} \\
VoxCPM & 1.52 & 0.743 & 2.46 & 0.675 & 14.29 & \textbf{0.704} & 2.651 \\
+SA-VAE & 1.63 & \underline{0.746} & 2.22 & 0.683 & 13.72 & \textbf{0.704} & 2.647 \\

\midrule

\multicolumn{8}{@{}l}{\textit{SemBridge Alignment--Anchoring}} \\
-Align,-Anchor & 1.58 & 0.745 & 2.43 & 0.685 & 16.87 & 0.696 & 2.622 \\
+Align & 1.51 & 0.735 & 2.30 & 0.678 & 15.94 & 0.697 & 2.620 \\
+Anchor & 1.21 & 0.742 & 2.18 & 0.683 & 13.97 & \underline{0.700} & \underline{2.659} \\
+Align,+Anchor & \textbf{1.01} & 0.743 & \textbf{1.87} & 0.687 & \textbf{11.87} & 0.697 & 2.648 \\

\midrule

\multicolumn{8}{@{}l}{\textit{Anchoring Depth $\lambda_{\mathrm{sem}}=0.1$}} \\
No Anchor & 1.51 & 0.735 & 2.30 & 0.678 & 15.94 & 0.697 & 2.620 \\
Anchor@18 & 1.13 & \textbf{0.747} & 2.12 & \underline{0.688} & 12.68 & 0.699 & 2.656 \\
Anchor@24 & \underline{1.04} & \textbf{0.747} & \underline{1.93} & \textbf{0.693} & \underline{12.01} & \textbf{0.704} & \textbf{2.683} \\
Anchor@32 & \textbf{1.01} & 0.743 & \textbf{1.87} & 0.687 & \textbf{11.87} & 0.697 & 2.648 \\

\midrule

\multicolumn{8}{@{}l}{\textit{Anchoring Strength} Anchor@32} \\
$\lambda_{\mathrm{sem}}=1.0$ & 1.64 & 0.681 & 5.11 & 0.591 & 24.31 & 0.632 & 2.588 \\
$\lambda_{\mathrm{sem}}=0.5$ & 1.61 & 0.705 & 3.64 & 0.623 & 16.25 & 0.663 & 2.614 \\
$\lambda_{\mathrm{sem}}=0.1$ & \textbf{1.01} & 0.743 & \textbf{1.87} & 0.687 & \textbf{11.87} & 0.697 & 2.648 \\
$\lambda_{\mathrm{sem}}=0.05$ & 1.13 & \underline{0.746} & 2.32 & 0.683 & 13.86 & \underline{0.700} & 2.652 \\

\bottomrule
\end{tabular}%
}
}
\caption{Ablation study of target-space semantic alignment and predictor-state semantic anchoring. All variants use the same 0.8B backbone, 100K hours of training data, and 300K updates. The default SemBridge configuration uses SA-VAE with semantic alignment and Anchor@32. UTMOS is reported on ZH-Hard for perceptual quality. Bold and underlined values indicate the globally best and second-best results, respectively. Tied results receive the same formatting.}
\label{tab:seed_tts_ablation}
\end{table}



\begin{table}[t]
\centering
{\footnotesize
\setlength{\tabcolsep}{2.5 pt}
\renewcommand{\arraystretch}{0.95}

\begin{tabular}{@{}lc|ccc@{}}
\toprule
\textbf{Model} & \textbf{Rate} & \textbf{PESQ} & \textbf{STOI} & \textbf{UTMOS} \\
\midrule
Reference Audio & -- & 4.64 & 1.00 & 4.09 \\

\midrule
\multicolumn{5}{l}{\textit{Discrete tokenizers}} \\
Encodec & 600 & 2.77 & 0.94 & 3.09 \\
Encodec & 150 & 1.56 & 0.85 & 1.58 \\
DAC & 400 & \textbf{4.01} & \underline{0.95} & 4.00 \\
SAC & 37.5 & 2.21 & 0.90 & \textbf{4.28} \\
Mimi & 100 & 2.26 & 0.91 & 3.63 \\
Mimi & 200 & 2.88 & 0.94 & 3.87 \\
WavTokenizer & 40 & 1.64 & 0.85 & 3.03 \\
X-codec2 & 50 & 2.43 & 0.92 & 4.13 \\
BiCodec & 50 & 2.51 & 0.92 & \underline{4.18} \\

\midrule
\multicolumn{5}{l}{\textit{Continuous tokenizer}} \\
VoxCPM-VAE & 25 & \underline{3.23} & \textbf{0.96} & 3.96 \\
VibeVoice-VAE & 7.5 & 2.98 & 0.75 & \underline{4.18} \\
\textbf{SA-VAE} & 25 & 2.99 & \textbf{0.96} & 3.92 \\
\midrule
w/o Alignment & 25 & 3.00 & \textbf{0.96} & 3.92 \\
\bottomrule
\end{tabular}
}
\caption{
Reconstruction quality of representative discrete
tokenizers and continuous tokenizer. Rate is
measured in frames or tokens per second; higher is better for
all metrics. Bold and underlined values indicate the globally
best and second-best results among tokenizers, respectively.
}
\label{tab:tokenizer_compact}
\end{table}


\begin{table*}[t]
\centering
\footnotesize
\setlength{\tabcolsep}{2.0pt}
\renewcommand{\arraystretch}{0.95}
\begin{tabular}{@{}lccccc|ccccc@{}}
\toprule
& \multicolumn{5}{c|}{\textbf{Mandarin}} & \multicolumn{5}{c}{\textbf{English}} \\
\cmidrule(lr){2-6}\cmidrule(lr){7-11}
\textbf{System} & CER (\%) $\downarrow$ & SIM $\uparrow$ & FFE (\%) $\downarrow$ & SingMOS $\uparrow$ & Sheet $\uparrow$ & WER (\%) $\downarrow$ & SIM $\uparrow$ & FFE (\%) $\downarrow$ & SingMOS $\uparrow$ & Sheet $\uparrow$ \\
\midrule
\multicolumn{11}{l}{\textit{Reference and reconstruction bounds}} \\
Reference Audio & 7.40 & -- & -- & 4.624 & 4.334 & 19.70 & -- & -- & 4.441 & 3.825 \\
\midrule
\multicolumn{11}{l}{\textit{Existing SVS systems}} \\
StyleSinger & 36.70 & 0.817 & 36.30 & 3.938 & 3.819 & -- & -- & -- & -- & -- \\
TCSinger & 27.00 & 0.855 & 31.50 & 3.977 & 3.843 & 41.00 & 0.879 & 21.10 & 3.662 & 3.482 \\
Vevo & 30.80 & 0.897 & \textbf{11.20} & 4.355 & 4.005 & 48.40 & \underline{0.924} & \textbf{8.80} & 4.321 & 3.699 \\
Ying-Music-Singer & 9.90 & 0.902 & 13.20 & 4.145 & 3.620 & -- & -- & -- & -- & -- \\
SoulX-Singer & \textbf{6.90} & \underline{0.905} & \underline{12.20} & 4.445 & 4.107 & \underline{14.90} & \textbf{0.926} & \underline{16.40} & 4.303 & 3.705 \\
\midrule
\multicolumn{11}{l}{\textit{Continuous-latent autoregressive system}} \\
\textbf{SemBridge}(ours) & \underline{8.32} & \textbf{0.906} & 16.70 & \textbf{4.585} & \textbf{4.315} & \textbf{14.77} & \textbf{0.926} & 25.89 & \textbf{4.444} & \textbf{3.994} \\
\midrule
w/o Anchoring & 9.18 & \underline{0.905} & 16.74 & \underline{4.560} & \underline{4.306} & 16.29 & \textbf{0.926} & 25.97 & \underline{4.432} & \underline{3.984} \\
\bottomrule
\end{tabular}%
\caption{Results on GMO-SVS. Lower CER/WER/FFE and higher SIM/SingMOS/Sheet are better. Bold and underlined values indicate the best and second-best synthesized results, respectively; tied results receive the same formatting.}
\label{tab:gmo_svs_main}
\end{table*}

\section{Experiments}
\label{sec:experiments}

\subsection{Experimental Setup}
\label{sec:experimental_setup}

\paragraph{Datasets.}
We evaluate \methodname{} primarily on zero-shot text-to-speech (TTS), with
score-conditioned singing voice synthesis (SVS) as a cross-task extension.
SA-VAE is trained separately on 25K hours of audio, including 20K hours of
speech sampled from public corpora and a 5K-hour subset of our internal
singing data. All matched TTS models and component ablations are trained on
the 100K-hour bilingual open-source VoxBox corpus~\cite{wang2025spark}. For
joint TTS and SVS training, we augment VoxBox with the full 20K hours of
internal singing data, resulting in a 120K-hour multitask corpus. We use the
100K-hour models for controlled component attribution and the 120K-hour model
for system-level comparison and SVS evaluation. Further details on corpus
composition and preprocessing are provided in the supplementary material.

\paragraph{Model Architecture.}
\methodname{} consists of a continuous-latent autoregressive backbone, a
frozen Semantic-Aligned Acoustic VAE (SA-VAE), and training-time semantic
supervision modules. The causal LM contains 32 Transformer blocks with a
hidden size of 1,024 and 16 attention heads. PatchEnc and LocDiT each contain
8 blocks with the same hidden size and number of heads. SA-VAE is a causal
convolutional Gaussian VAE that encodes 44.1~kHz waveforms into
64-dimensional continuous latent frames at 25~Hz without vector quantization.
The frozen semantic tokenizer operates at 12.5~Hz with a vocabulary of
16,384 tokens. During Stage~I, a lightweight projection aligns SA-VAE
acoustic patches with token-level semantic embeddings. During Stage~II, the
corresponding discrete token labels supervise autoregressive LM states. We
denote a semantic head attached after Transformer block \(k\) as
Anchor@\(k\).

Excluding the frozen SA-VAE, the continuous generation backbone contains
800.15M trainable parameters. SA-VAE contains 86.48M parameters, while the
alignment projection and semantic anchoring module add 2.43M and 16.79M
parameters, respectively. The alignment projection and semantic head are used
only during training and are removed at inference.

\paragraph{Training Details.}
We train SA-VAE independently for 300K updates on 3.0-second audio segments
with a batch size of 48 using 8 NVIDIA H20 GPUs. We then train
\methodname{} for 300K updates using 16 NVIDIA H20 GPUs and a global batch
size of 4,096 acoustic-patch frames. We use AdamW
~\cite{DBLP:conf/iclr/LoshchilovH19} with bfloat16 precision. The learning
rate is linearly increased to \(1\times10^{-4}\) over the first 5K updates
and then decayed with a cosine schedule. The flow-matching and stop-prediction
losses are assigned unit weights, and the semantic anchoring weight is set to
\(\lambda_{\mathrm{sem}}=0.1\). Inference uses 10 function evaluations
(\(\mathrm{NFE}=10\)), a classifier-free guidance scale of 2.0
~\cite{ho2022classifier}, and a sampling temperature of 1.0.

\paragraph{Evaluation Metrics.}
We evaluate zero-shot TTS on Seed-TTS-Eval
~\cite{anastassiou2024seedtts} and CV3-Eval
~\cite{du2025cosyvoice3}. Content fidelity is measured by character error
rate (CER) for Chinese using Paraformer~\cite{gao2022paraformer} and word
error rate (WER) for English using Whisper-large-v3
~\cite{radford2023whisper}. Speaker similarity (SIM) is computed as the
cosine similarity between WavLM-based speaker embeddings
~\cite{chen2022wavlm}, and UTMOS~\cite{saeki2022utmos} is used as an
automatic estimate of perceptual quality. We evaluate SA-VAE reconstruction
using PESQ~\cite{DBLP:conf/icassp/RixBHH01}, STOI
~\cite{DBLP:journals/taslp/TaalHHJ11}, and UTMOS. For SVS, we follow the
GMO-SVS evaluation protocol~\cite{qian2026soulxsinger} and report
CER/WER, SIM, F0 frame error (FFE), SingMOS-Pro
~\cite{tang2025singmos}, and Sheet scores.

\paragraph{Baselines.}
For zero-shot TTS, we compare \methodname{} with representative systems from
several speech generation paradigms. Non-autoregressive baselines include
F5-TTS~\cite{chen2024f5} and MaskGCT~\cite{wang2024maskgct}, which use
flow-matching and masked-generation formulations, respectively. Discrete-token
autoregressive systems include the CosyVoice series
~\cite{cosyvoice,cosyvoice2,du2025cosyvoice3}, Spark-TTS
~\cite{wang2025spark}, FireRedTTS~\cite{guo2024fireredtts}, HiggsAudio-v2\cite{bosonai_higgs_audio_tts_v3_2026},and
IndexTTS2~\cite{zhou2025indextts2}. We also compare with continuous-latent or
hybrid autoregressive systems, including DiTAR~\cite{jia2025ditar},
VoxCPM~\cite{zhou2025voxcpm}, VoxCPM2
~\cite{DBLP:journals/corr/abs-2606-06928}, VibeVoice
~\cite{peng2026vibevoice}, MELA-TTS~\cite{an2025melatts}, and
SemaVoice~\cite{wang2026semavoice}.

For score-conditioned svs, we compare with representative recent systems on
GMO-SVS, including StyleSinger~\cite{DBLP:conf/aaai/ZhangHLHXCDHZ24},
TCSinger~\cite{DBLP:conf/emnlp/01260LPHHWZ24},
Vevo~\cite{DBLP:conf/iclr/ZhangZPTMLHLWCH25},
Ying-Music-Singer~\cite{DBLP:journals/corr/abs-2512-04779}, and
SoulX-Singer~\cite{qian2026soulxsinger}. Baseline results are taken from the
original publications or obtained using official checkpoints when available.
The supplementary material reports the model scale, training data, evaluation
source, and implementation details for each comparison.

\subsection{Experimental Results}
\label{sec:experimental_results}

\paragraph{Main Results on Zero-Shot TTS.}
Table~\ref{tab:speech_generation_main} reports zero-shot TTS results on
Seed-TTS-Eval and CV3-EVAL. The system-level \methodname{} model, trained on
the 120K-hour multitask corpus, achieves consistently low content error across
languages and evaluation conditions while maintaining competitive speaker
similarity. On Seed-TTS-Eval, it obtains a CER of 0.95 on Chinese, a WER of
1.81 on English, and a CER of 9.79 on the Chinese hard subset. On CV3-EVAL,
it achieves a CER of 3.34 on Chinese, a WER of 4.22 on English, a CER of
10.58 on the Chinese hard subset, and a WER of 6.35 on the English hard
subset. Compared with MELA-TTS, which achieves competitive content accuracy
but lower speaker similarity, particularly on English, \methodname{} provides
a more balanced content--speaker trade-off. This pattern is consistent with
token-level classification providing a more focused semantic regularizer than
full continuous feature matching, though the cross-system comparison does not
establish causality.

\subsubsection{Controlled Analysis of Semantic-Token Anchoring}
\label{sec:anchoring_analysis}

\paragraph{Acoustic Representation Control.}
Table~\ref{tab:seed_tts_ablation} reports controlled experiments using the
same 0.8B model scale, 100K hours of training data, and 300K updates. For the
continuous-AR baseline, replacing the original acoustic representation with
SA-VAE reduces English WER from 2.46 to 2.22 and ZH-Hard CER from 14.29 to
13.72, while Chinese CER changes from 1.52 to 1.63. Although gains vary across
subsets, the results show that acoustic representation affects content modeling
and motivate the subsequent component analysis under a fixed SemBridge
backbone.

\paragraph{Alignment and Anchoring Attribution.}
Within the SemBridge backbone, the model without target alignment or state
anchoring obtains errors of 1.58, 2.43, and 16.87 on ZH, EN, and ZH-Hard,
respectively. Target-space alignment alone reduces these errors to 1.51, 2.30,
and 15.94, whereas semantic-token anchoring produces larger reductions to
1.21, 2.18, and 13.97. Combining the two objectives gives the lowest errors
of 1.01, 1.87, and 11.87. These results indicate that state-level
semantic-token anchoring drives most content-fidelity gains, while target-space
alignment provides additional complementary gains.

\paragraph{Anchoring Depth.}
We vary anchoring depth by attaching the semantic prediction head after
different Transformer blocks. Anchor@32 yields the lowest content errors,
with a CER of 1.01 on ZH, a WER of 1.87 on EN, and a CER of 11.87 on
ZH-Hard. Anchor@24 increases these errors only slightly to 1.04, 1.93, and
12.01, while improving SIM from 0.743, 0.687, and 0.697 to 0.747, 0.693,
and 0.704, respectively. It also raises ZH-Hard UTMOS from 2.648 to 2.683.
We therefore use Anchor@24 for the system-level model because it better
balances content accuracy, speaker similarity, and perceptual quality.
Overall, anchoring depth trades off content and quality metrics.

\paragraph{Anchoring Strength.}
We vary \(\lambda_{\mathrm{sem}}\) to assess semantic supervision strength.
Increasing \(\lambda_{\mathrm{sem}}\) from 0.1 to 0.5 worsens English WER from
1.87 to 3.64 and reduces SIM from 0.687 to 0.623. Setting
\(\lambda_{\mathrm{sem}}=1.0\) further degrades content, similarity, and UTMOS
metrics. A smaller weight of 0.05 preserves speaker similarity and perceptual
quality but gives higher content errors than 0.1. We therefore use
\(\lambda_{\mathrm{sem}}=0.1\), which minimizes content errors while
maintaining comparable synthesis quality. These results suggest that semantic
anchoring should regularize autoregressive state learning without dominating
the continuous generation objective.

\subsubsection{SA-VAE Reconstruction under Semantic Alignment}
\label{sec:representation_analysis}

Table~\ref{tab:tokenizer_compact} compares SA-VAE reconstruction with
representative discrete and continuous speech representations. At 25~Hz,
SA-VAE achieves a PESQ of 2.99, a STOI of 0.96, and a UTMOS score of 3.92.
Its unaligned counterpart obtains nearly identical scores of 3.00, 0.96, and
3.92, respectively, showing that semantic alignment does not noticeably
degrade acoustic reconstruction. Despite different representation rates and
architectures, SA-VAE remains competitive with both continuous and discrete
tokenizers. These results support its role as a complementary target-space
component that introduces semantic alignment without imposing a discrete
bottleneck or sacrificing reconstruction quality.

\paragraph{Cross-Task Transfer to Score-Conditioned SVS.}

Table~\ref{tab:gmo_svs_main} evaluates semantic-token anchoring on
score-conditioned SVS. Compared with the model without anchoring, SemBridge
reduces Mandarin CER from 9.18 to 8.32 and English WER from 16.29 to 14.77,
while speaker similarity remains essentially unchanged. SingMOS and Sheet
scores also improve slightly in both languages, showing that improved lyric
intelligibility does not compromise perceived singing quality or
expressiveness. SemBridge further achieves the highest synthesized SingMOS and
Sheet scores among the compared systems. Consistent gains across Mandarin and
English suggest that the semantic constraint remains effective when linguistic
prediction is coupled with score-derived pitch and timing conditions. Its FFE
results are not the best, indicating that semantic-token anchoring mainly
improves lyric modeling and perceptual quality rather than pitch accuracy.
Overall, the proposed supervision transfers beyond TTS without requiring a
task-specific semantic branch.

\section{Conclusion}

In this paper, we introduced \methodname{}, a training-only semantic-token
anchoring framework for continuous-latent autoregressive speech generation.
\methodname{} directly supervises autoregressive LM states with discrete
semantic-token labels, while a complementary Semantic-Aligned Acoustic VAE
aligns continuous acoustic targets with embeddings from the same frozen
semantic tokenizer. This shared semantic reference strengthens state-level
linguistic modeling and target-space semantic correspondence while preserving
continuous-only autoregressive inference. Experiments on zero-shot TTS and
score-conditioned SVS demonstrate that \methodname{} consistently improves
content fidelity while maintaining competitive speaker similarity and
perceptual quality. Controlled ablations further confirm the complementary
roles of latent alignment and state anchoring and reveal a trade-off effect
of anchoring depth. Overall, semantic-token anchoring provides an effective
training strategy for improving linguistic modeling in continuous-latent
autoregressive speech generation.

\bibliography{aaai2027}

\FloatBarrier
\raggedbottom

\providecommand{\methodname}{SemBridge}
\providecommand{\tbd}[1]{\textbf{[TBD: #1]}}

\appendix

\section{Training Data}
\label{sec:supp_data}

SA-VAE and the \methodname{} are trained in separate stages using
 counted training corpora. SA-VAE is trained on 19,906 hours of
Mandarin and English speech together with a 5,000-hour subset of singing data,
yielding 24,906 hours of audio in total. The matched TTS configurations are
trained on 100K hours of bilingual VoxBox speech. The Joint TTS--SVS model
additionally uses the complete 20K-hour score-aligned singing collection,
resulting in a 120K-hour corpus.Table~\ref{tab:supp_training_data} summarizes the data used at each training stage.

\paragraph{SA-VAE Training Data.}
Panel~(a) of Table~\ref{tab:supp_training_data} lists the composition of the
SA-VAE training corpus. The Mandarin speech data consist of Emilia-ZH,
WenetSpeech4TTS, and in-house paralinguistic speech. The English speech data
consist of LibriSpeech, the small and medium LibriHeavy subsets, Emilia-EN,
MLS-EN, and in-house speech. Together, these sources provide 19,906 hours of
bilingual speech, including 19,172 hours from public corpora and 734 hours from
in-house data.

We further sample 5,000 hours from the internal singing corpus.
Stage~I uses only the waveforms and does not use the corresponding lyric or
score annotations. The complete SA-VAE training set therefore contains
24,906 hours of audio, referred to as approximately 25K hours.

\paragraph{\methodname{} Training Data.}
All controlled TTS experiments, including the component, anchoring-depth, and
loss-weight studies, use the same 100K-hour Mandarin--English VoxBox speech
corpus. This setting keeps the generator training data fixed across all
controlled component comparisons.

The Joint TTS--SVS model augments the same speech corpus with the complete
20K-hour internal score-aligned singing collection, resulting in 120K hours of
generator training data. The 5K-hour singing subset used for SA-VAE training is
drawn from this 20K-hour collection. Because SA-VAE is trained separately and
then frozen, its 24,906-hour corpus is not added to the 120K-hour generator
data count. Results obtained with the Joint model are therefore treated as
system-level comparisons rather than controlled component ablations.

\begin{table}[t]
\centering
\footnotesize
\setlength{\tabcolsep}{2.2pt}
\renewcommand{\arraystretch}{0.96}

\begin{tabularx}{\columnwidth}{
@{}>{\raggedright\arraybackslash}X
l
r
r
r@{}
}
\toprule
\multicolumn{5}{@{}l}{\textbf{(a) SA-VAE training data}} \\
\midrule
\textbf{Dataset}
& \textbf{Group}
& \textbf{\#Utt.}
& \textbf{Hours}
& \textbf{Avg.~s} \\
\midrule
Emilia-ZH       & ZH & 3.0M & 6{,}712 & 8.05 \\
WenetSpeech4TTS & ZH & 2.0M & 2{,}754 & 4.96 \\
In-house Data   & ZH & 0.2M & 525     & 7.82 \\
\midrule
LibriSpeech     & EN & 0.3M & 961     & 12.30 \\
LibriHeavy      & EN & 1.2M & 5{,}042 & 14.85 \\
Emilia-EN       & EN & 1.0M & 2{,}481 & 8.93 \\
MLS-EN          & EN & 0.3M & 1{,}222 & 14.66 \\
In-house Data   & EN & 0.1M & 209     & 6.42 \\
\midrule
\textbf{Speech subtotal}
& ZH/EN
& \textbf{8.2M}
& \textbf{19{,}906}
& \textbf{8.78} \\
In-house Singing
& Singing
& 2.2M
& 5{,}000
& 8.18 \\
\midrule
\textbf{SA-VAE total}
& All
& \textbf{10.4M}
& \textbf{24{,}906}
& \textbf{8.62} \\
\bottomrule
\end{tabularx}

\vspace{0.55em}

\begin{tabularx}{\columnwidth}{
@{}l
>{\raggedright\arraybackslash}X
>{\raggedright\arraybackslash}X
r@{}
}
\toprule
\multicolumn{4}{@{}l}{\textbf{(b) \methodname{} generator training data}} \\
\midrule
\textbf{Setting}
& \textbf{Speech pool}
& \textbf{Singing pool}
& \textbf{Total (h)} \\
\midrule
Matched TTS
& VoxBox (100K h)
& --
& 100K \\
Joint TTS--SVS
& VoxBox (100K h)
& In-house (20K h)
& 120K \\
\bottomrule
\end{tabularx}

\caption{Training data for SA-VAE and \methodname{}.}
\label{tab:supp_training_data}

\end{table}

\section{Experimental Configuration}
\label{sec:supp_setting}

\begin{table*}[t]
\centering
\small

\begin{tabularx}{\textwidth}{@{}lXlX@{}}
\toprule
\textbf{Setting} & \textbf{Value} & \textbf{Setting} & \textbf{Value} \\
\midrule
Waveform rate & 44.1~kHz & Formulation & Causal convolutional Gaussian VAE \\
Encoder downsampling factors & [2, 3, 6, 7, 7] & Decoder upsampling factors & [7, 7, 6, 3, 2] \\
Hop size & 1,764 samples & Acoustic latent rate & 25~Hz \\
Latent dimension & 64 & Patch shape and rate & \(2\times64\) at 12.5~Hz \\
Semantic vocabulary size & 16,384 & Semantic-token rate & 12.5~Hz \\
SA-VAE parameters & 86.48M & Alignment projection parameters & 2.43M \\
Acoustic quantization & None & Generator modeling unit & Continuous acoustic latent patch \\
\bottomrule
\end{tabularx}
\caption{SA-VAE architecture and semantic-alignment configuration.}
\label{tab:supp_savae_architecture}
\end{table*}

\begin{table*}[t]
\centering
\small

\begin{tabular}{@{}lcccccl@{}}
\toprule
\textbf{Component} & \textbf{Blocks/Layers} & \textbf{Width} & \textbf{Heads} & \textbf{Rate} & \textbf{Stage II} & \textbf{Role} \\
\midrule
PatchEnc & 8 & 1,024 & 16 & 12.5~Hz & Trainable & Acoustic prompt and history encoding \\
Causal LM & 32 & 1,024 & 16 & 12.5~Hz & Trainable & Causal next-patch state modeling \\
LocDiT & 8 & 1,024 & 16 & 12.5~Hz & Trainable & Flow-matching patch generation \\
Stop predictor & 2-layer MLP & 1,024 & -- & 12.5~Hz & Trainable & Stop/continue classification \\
Semantic head & Linear & 1,024 & -- & 12.5~Hz & Trainable & 16,384-unit semantic-token classification \\
SA-VAE decoder & Causal conv. & -- & -- & 25Hz & Frozen & Waveform reconstruction \\
\bottomrule
\end{tabular}
\caption{Architecture of \methodname{}. }
\label{tab:supp_SemBridge_architecture}
\end{table*}

\subsection{Stage I: Semantic-Aligned Acoustic VAE}

\label{sec:supp_savae_method}

SA-VAE is a causal convolutional Gaussian VAE that operates on 44.1-kHz
waveforms. The encoder produces 64-dimensional continuous latent frames at
25~Hz without vector quantization. For each training segment, a latent
sequence is sampled from the diagonal Gaussian posterior through
reparameterization. The same sample is passed to both the waveform decoder and
the semantic-alignment projection. Every two adjacent latent frames are
concatenated into a \(2\times64\) acoustic patch at 12.5~Hz. The resulting
patch is projected into the embedding space of the frozen semantic tokenizer
and aligned with the corresponding stop-gradient semantic embedding.

The patch construction and semantic-alignment objective are defined in
Eqs.~(2), (14), and (15) of the main paper. The complete loss
objective used in our implementation is

\begin{equation}
\begin{aligned}
\mathcal L_{\mathrm{SA\text{-}VAE}}
={}&
10\mathcal L_{\mathrm{MR\text{-}STFT}}
+15\mathcal L_{\mathrm{Mel}}
+0.01\mathcal L_{\mathrm{KL}} \\
&+\mathcal L_{\mathrm{adv}}
+1.5\mathcal L_{\mathrm{feat}}
+100\mathcal L_{\mathrm{align}}.
\end{aligned}
\label{eq:supp_savae_objective}
\end{equation}

The alignment loss is averaged over valid acoustic-patch positions and
combines a dimension-normalized \(\ell_1\) distance with
\(\operatorname{softplus}(-\cos(\cdot,\cdot))\). Gradients are not propagated
through the semantic embeddings produced by the frozen tokenizer. In the
alignment ablation, only \(\mathcal L_{\mathrm{align}}\) is removed; the
encoder--decoder architecture and all acoustic reconstruction losses remain
unchanged.

The product of the encoder downsampling factors is
\(2\times3\times6\times7\times7=1764\), which yields exactly 25 latent frames
per second at 44.1~kHz. The 12.5-Hz acoustic-patch and semantic-token sequences
use the same valid-position mask. Incomplete final patches and padded
positions are excluded from \(\mathcal L_{\mathrm{align}}\).

The adversarial objective uses a multi-period discriminator and a multi-scale
sub-band CQT discriminator following the BigVGAN V2 configuration. The
adversarial and feature-matching losses are enabled after a 1,000-update
warm-up. SA-VAE is trained for 300K updates on 3-s segments with a global batch
size of 48 using eight NVIDIA H20 GPUs. The trained SA-VAE is frozen in all
subsequent \methodname{} generator experiments.

\subsection{Stage II: Continuous-Latent Autoregressive Generator}
\label{sec:supp_generator_method}

The \methodname{} generator follows a PatchEnc--LM--LocDiT architecture.
PatchEnc maps latent patches from the acoustic prompt and autoregressive
history to acoustic-context representations. A 32-block causal LM jointly
models these representations and a task-specific symbolic prefix. The prefix
contains text for TTS and additionally contains lyric, pitch, onset, and
duration sequences for score-conditioned SVS. An eight-block LocDiT performs
next-patch flow matching conditioned on the normalized final LM state and the
preceding SA-VAE latent patch. A two-layer
\(1024\!\rightarrow\!1024\!\rightarrow\!2\) MLP with a SiLU activation
predicts whether generation should continue or terminate.

Excluding the frozen SA-VAE, the continuous generation backbone contains
800.15M trainable parameters. The semantic classifier adds 16.79M parameters
during training and is removed at inference. SA-VAE contains 86.48M
parameters, and its 2.43M-parameter alignment projection is used only during
Stage~I.

\subsection{Training Configuration}
\label{sec:supp_optimization}

SA-VAE is optimized independently and frozen before generator training.
Unless otherwise stated, all matched \methodname{} configurations use the same
model architecture, optimization schedule, random seed, and 300K-update
training budget.

\begin{table*}[t]
\centering
\small

\begin{tabularx}{\textwidth}{@{}lXlX@{}}
\toprule
\multicolumn{2}{c}{\textbf{SA-VAE}} &
\multicolumn{2}{c}{\textbf{\methodname{} Generator}} \\
\cmidrule(lr){1-2}\cmidrule(lr){3-4}
\textbf{Setting} & \textbf{Value} & \textbf{Setting} & \textbf{Value} \\
\midrule
Training budget & 300K updates & Training budget & 300K updates \\
Training input & 3s waveforms segments at 44.1~kHz & Modeling unit & 12.5-Hz continuous acoustic patches \\
Global batch size & 48 segments & Global batch size & 4,096 acoustic-patch frames \\
Hardware & \(8\times\) NVIDIA H20 & Hardware & \(16\times\) NVIDIA H20 98~GB \\
Precision & fp32 & Precision & bfloat16 \\
Optimizer & AdamW & Optimizer & AdamW \\
Learning rate & \(1\times10^{-4}\) & Learning-rate schedule & Peak \(1\times10^{-4}\); 5K warm-up; cosine decay \\
Loss weights & MR-STFT 10; mel 15; KL 0.01; alignment 100; adversarial 1.0; feature matching 1.5 & Loss weights & Flow matching 1.0; semantic anchoring 0.1; stop prediction 1.0 \\
Global seed & 42 & Global seed & 42 \\
\bottomrule
\end{tabularx}
\caption{Training configurations of SA-VAE and the \methodname{} generator.}
\label{tab:supp_training_configuration}
\end{table*}

The global seed is fixed to 42 for Python, NumPy, PyTorch, and CUDA in all
reported experiments.

\begin{table*}[t]
\centering
\footnotesize
\setlength{\tabcolsep}{2.2pt}
\renewcommand{\arraystretch}{0.95}
\resizebox{0.98\textwidth}{!}{%
\begin{tabular}{@{}llcc|cccccc@{}}
\toprule
\textbf{Model}
& \textbf{Alignment Loss}
& \(\boldsymbol{\beta}_{\mathrm{KL}}\)
& \textbf{Frame Rate (Hz)}
& \textbf{PESQ-WB} $\uparrow$
& \textbf{PESQ-NB} $\uparrow$
& \textbf{STOI} $\uparrow$
& \textbf{UTMOS} $\uparrow$
& \textbf{SIM} $\uparrow$
& \textbf{Recon.\ WER} $\downarrow$ \\
\midrule
Vanilla VAE & None & \(5\times10^{-5}\) & 25 & \textbf{3.186} & \textbf{3.562} & \textbf{0.963} & 3.888 & \textbf{0.942} & 2.312 \\
Vanilla VAE & None & \(0.01\) & 25 & 2.996 & 3.421 & \underline{0.960} & 3.890 & 0.926 & \underline{2.292} \\
SA-VAE & \(\mathcal{L}_{\mathrm{MSE}}\) & \(0.01\) & 25 & \underline{3.071} & \underline{3.505} & 0.959 & \underline{3.914} & \underline{0.939} & 2.317 \\
SA-VAE & \(\mathcal{L}_{\cos}\) & \(0.01\) & 25 & 3.063 & 3.501 & 0.957 & 3.884 & 0.932 & 2.300 \\
\textbf{SA-VAE} & \(\mathcal{L}_{\cos}+\mathcal{L}_{1}\) & \(0.01\) & 25 & 2.988 & 3.449 & 0.958 & \textbf{3.923} & 0.930 & \textbf{2.268} \\
\bottomrule
\end{tabular}%
}

\caption{
Reconstruction ablation over KL weights and semantic-alignment losses.
Best and second-best results are shown in \textbf{bold} and
\underline{underlined}, respectively.
}
\label{tab:supp_savae_reconstruction_ablation}
\end{table*}

\section{Evaluation Metrics}
\label{sec:supp_evaluation}

\paragraph{Zero-shot TTS.}
Seed-TTS-Eval contains 2,020 Mandarin examples from DiDiSpeech and 1,088
English examples from Common Voice, together with a Mandarin hard subset.
For both Seed-TTS-Eval and CV3-EVAL, all locally evaluated systems use the
prompt--target pairs, transcripts, and data partitions provided by the
official benchmark manifests.

When a system returns the acoustic prompt and generated continuation as a
single waveform, the prompt region is removed before metric computation.
Speaker similarity is computed between the generated target region and the
corresponding prompt. The same output-trimming and failure-handling rules are
applied to all locally evaluated systems. No utterance is excluded according
to its metric value, and results are aggregated separately for each language
and benchmark subset.

\paragraph{Score-conditioned SVS.}
We evaluate score-conditioned SVS on GMO-SVS and SoulX-Singer-Eval following
the SoulX-Singer evaluation protocol. GMO-SVS contains 802 samples collected
from GTSinger, M4Singer, and OpenCpop. The first sentence of each song is used
as the acoustic prompt, and the remaining sentences are used as synthesis
targets. We evaluate the original-lyric score-conditioned setting rather than
the rewritten-lyric editing setting.

SoulX-Singer-Eval contains 100 prompt segments from 50 unseen singers,
including 25 Mandarin and 25 English singers, with two segments per singer.
The Mandarin prompts are collected from professional and amateur singers, and
the English prompts are drawn from the Mixing Secrets dataset. Target lyrics
and melodies are selected from 15 Mandarin and 15 English tracks in GMO-SVS
and paired with manually verified note-level annotations. We use the official
same-language Mandarin and English evaluation partitions.
\paragraph{Metric Implementation.}
Content, pitch, and perceptual-quality metrics are computed only on the
generated target region. The acoustic prompt is used only as the reference for
speaker similarity. Mandarin recognition error is reported as CER, although
the original SoulX-Singer protocol uses the term WER for both Mandarin and
English recognition results. On GMO-SVS, an F0 frame is counted as incorrect
when the relative deviation from the reference F0 exceeds 20\%. Following the
official SoulX-Singer-Eval protocol, FFE is not reported on that benchmark.
CER, WER, and FFE are multiplied by 100 and reported as percentages.


\begin{table}[t]
\centering
{\footnotesize
\setlength{\tabcolsep}{1.4pt}
\renewcommand{\arraystretch}{0.95}
\resizebox{\columnwidth}{!}{%
\begin{tabular}{@{}l|cc|cc|cc@{}}
\toprule
\textbf{VAE Configuration}
& \multicolumn{2}{c|}{\textbf{ZH}}
& \multicolumn{2}{c|}{\textbf{EN}}
& \multicolumn{2}{c}{\textbf{ZH-Hard}} \\
& CER $\downarrow$
& SIM $\uparrow$
& WER $\downarrow$
& SIM $\uparrow$
& CER $\downarrow$
& SIM $\uparrow$ \\
\midrule
Vanilla VAE (\(\beta_{\mathrm{KL}}=5\times10^{-5}\)) & 1.99 & 0.701 & 3.43 & 0.622 & 17.27 & \textbf{0.664} \\
Vanilla VAE (\(\beta_{\mathrm{KL}}=0.01\)) & 1.72 & 0.695 & 3.05 & 0.617 & \underline{15.95} & 0.654 \\
SA-VAE (\(\mathcal{L}_{\mathrm{MSE}}\)) & 1.48 & 0.7022 & 3.09 & \underline{0.6293} & 18.61 & 0.655 \\
SA-VAE (\(\mathcal{L}_{\cos}\)) & \textbf{1.45} & \underline{0.709} & \underline{2.96} & 0.628 & 16.16 & \underline{0.663} \\
SA-VAE (\(\mathcal{L}_{\cos}+\mathcal{L}_{1}\); default) & \underline{1.47} & \textbf{0.711} & \textbf{2.81} & \textbf{0.631} & \textbf{15.94} & \textbf{0.664} \\
\bottomrule
\end{tabular}%
}
}
\caption{
Downstream comparison of VAE configurations. Best and second-best results are
shown in \textbf{bold} and \underline{underlined}, respectively.
}
\label{tab:supp_savae_downstream_ablation}
\end{table}

\section{Additional Experimental Results} 
\label{sec:supp_additional_results} 
This section distinguishes controlled diagnostic studies from final-checkpoint
evaluations. Unless otherwise stated, ablations of the SA-VAE configuration and
state-supervision objective use the 100K-hour speech-only corpus, a shared 0.8B
backbone architecture, and a matched budget of 100K updates. Convergence analyses
and final-checkpoint comparisons follow the full 300K-update schedule and are
explicitly identified. Inference settings are fixed across comparisons, except
when classifier-free guidance (CFG) is varied.

\subsection{SA-VAE Representation Analysis}

\label{sec:supp_results}

\paragraph{Ablations on Alignment Loss and KL Weight.}
All configurations use 25-Hz continuous latents and share the same
encoder--decoder architecture, training data, optimization budget, and
evaluation protocol. The first two rows isolate the effect of
\(\beta_{\mathrm{KL}}\) without semantic alignment, whereas the remaining rows
compare alignment losses at \(\beta_{\mathrm{KL}}=0.01\) against the matched
Vanilla VAE. Reducing the KL weight yields the strongest PESQ, STOI, and SIM,
but the alignment objectives reveal a trade-off between signal fidelity and
semantic reconstruction. Although \(\mathcal{L}_{\mathrm{MSE}}\) better
preserves PESQ and SIM, \(\mathcal{L}_{\cos}+\mathcal{L}_{1}\) achieves the
highest UTMOS (3.923) and lowest reconstruction WER (2.268), improving over the
matched baseline at 3.890 and 2.292, respectively. We therefore adopt
\(\mathcal{L}_{\cos}+\mathcal{L}_{1}\) for its favorable balance of signal
fidelity, perceptual quality, and linguistic intelligibility.

\paragraph{SA-VAE Provides the Best Downstream Balance.}
All variants use the same 0.8B backbone, 100K-hour speech-only corpus,
100K-update training budget, GLM-4 semantic-token targets, and Anchor@32
configuration, thereby isolating the effect of the pretrained and subsequently
frozen VAE. Among the Vanilla VAE configurations, increasing
\(\beta_{\mathrm{KL}}\) from \(5\times10^{-5}\) to \(0.01\) improves content
accuracy on all three evaluation sets, but consistently reduces SIM. At the
matched \(\beta_{\mathrm{KL}}=0.01\), the downstream results are sensitive to
the alignment objective: \(\mathcal{L}_{\mathrm{MSE}}\) improves ZH CER but
degrades EN WER and ZH-Hard CER relative to the Vanilla VAE, indicating an
inconsistent transfer benefit. In contrast,
\(\mathcal{L}_{\cos}+\mathcal{L}_{1}\) obtains the best or tied-best result on
five of the six metrics. Relative to the matched Vanilla VAE, it reduces ZH CER
by \(14.5\%\) and EN WER by \(7.9\%\), leaves ZH-Hard CER effectively unchanged
at 15.94 versus 15.95, and increases SIM by 0.016, 0.014, and 0.010 on ZH, EN,
and ZH-Hard, respectively. Although \(\mathcal{L}_{\cos}\) achieves a slightly
lower ZH CER than the default configuration (1.45 versus 1.47), the combined
objective performs better on each of the remaining five metrics. Together with
its lowest reconstruction WER in
Table~\ref{tab:supp_savae_reconstruction_ablation}, these results support
\(\mathcal{L}_{\cos}+\mathcal{L}_{1}\) as the default for balancing content
accuracy and speaker similarity.

\paragraph{Representation-Space Analysis of SA-VAE}
\label{sec:supp_savae_representation}

\begin{figure}[t]
  \centering
  \includegraphics[width=\linewidth]{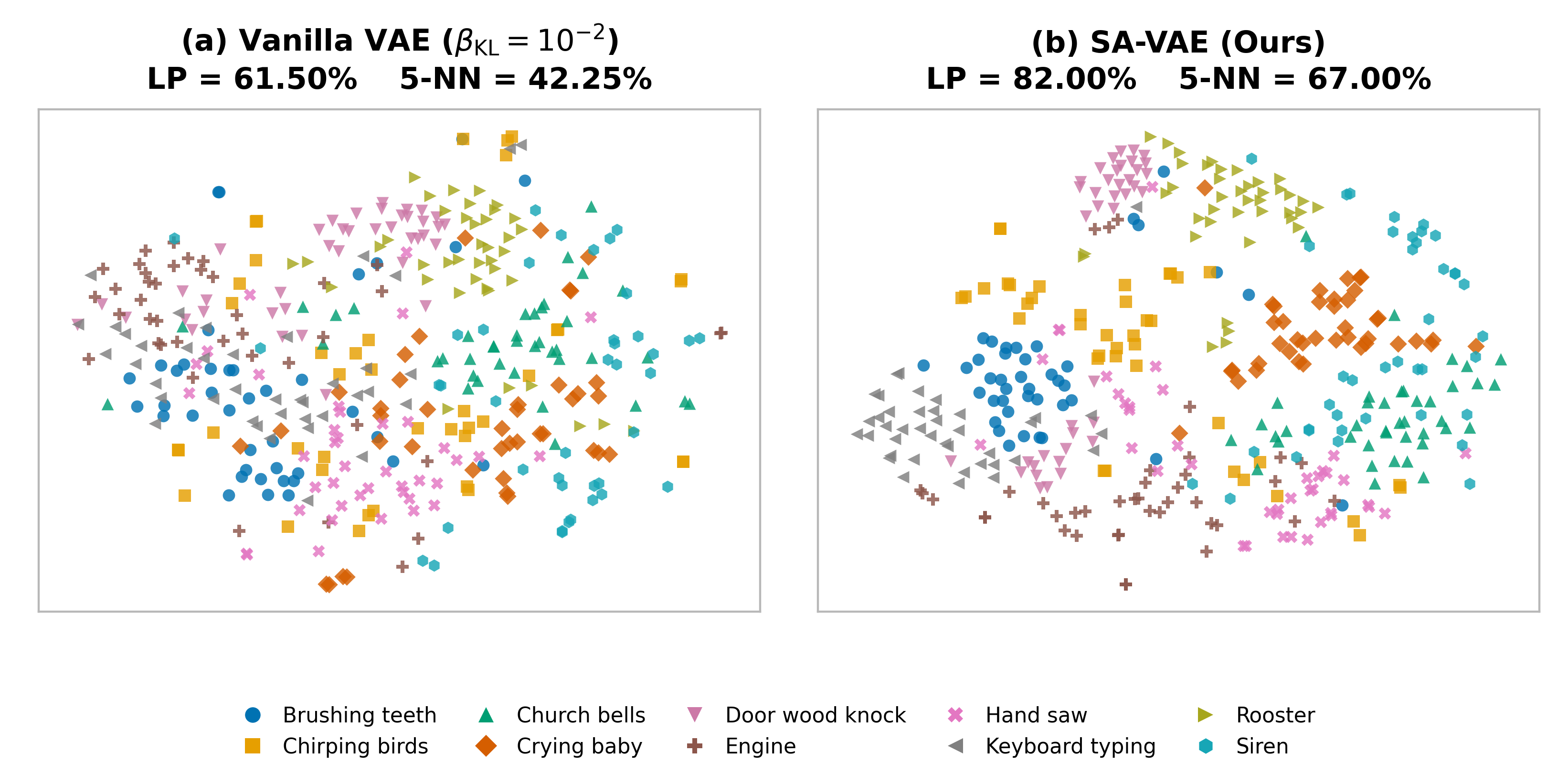}
  \caption{
  Representation-space comparison between a Vanilla VAE
  ($\beta_{\mathrm{KL}}=10^{-2}$) and the proposed SA-VAE on ten
  ESC-50 sound categories. The t-SNE projections are obtained from
  temporally pooled 64-dimensional sampled latents $\mathbf{z}$.
  Linear-probe (LP) and 5-nearest-neighbor (5-NN) accuracies are
  computed in the original standardized latent space rather than in
  the two-dimensional t-SNE space. Higher values indicate stronger
  linear class separability and more consistent local neighborhood
  structure, respectively.
  }
  \label{fig:savae_representation_space}
\end{figure}

Figure~\ref{fig:savae_representation_space} compares the latent-space organization of the Vanilla VAE and SA-VAE under the same evaluation protocol. The Vanilla VAE representations exhibit substantial overlap across sound
categories, indicating that reconstruction and KL regularization alone do not explicitly organize the continuous latent space according to category-relevant content. In contrast, SA-VAE produces visibly more compact and better
separated clusters, while preserving a continuous 64-dimensional acoustic representation. This improvement is also reflected by the quantitative measurements: SA-VAE increases the linear-probe accuracy from $61.50\%$ to $82.00\%$,
corresponding to an absolute gain of $20.50$ percentage points, and improves the 5-NN accuracy from $42.25\%$ to $67.00\%$, an absolute gain of $24.75$ percentage points. The higher LP accuracy shows that category-relevant
information becomes more directly accessible through a linear classifier, whereas the higher 5-NN accuracy indicates that samples from the same category form more consistent local neighborhoods. Since the category labels are used
only for evaluation and are not provided as training targets, these results suggest that alignment with the frozen semantic tokenizer regularizes the acoustic latent space toward a more structured representation without introducing
categorical supervision. The t-SNE projections are used only as qualitative illustrations; the quantitative conclusion is supported by LP and 5-NN measurements computed before dimensionality reduction.

\subsection{Analysis of Semantic-Token Anchoring}
\label{sec:supp_anchoring_analysis}


\begin{table}[t]
\centering
{\footnotesize
\setlength{\tabcolsep}{0.8pt}
\renewcommand{\arraystretch}{0.95}
\resizebox{\columnwidth}{!}{%
\begin{tabular}{@{}llc|cc|cc|cc@{}}
\toprule
\textbf{Target} & \textbf{Loss} & \(\lambda_{\mathrm{sem}}\)
& \multicolumn{2}{c|}{\textbf{ZH}}
& \multicolumn{2}{c|}{\textbf{EN}}
& \multicolumn{2}{c}{\textbf{ZH-Hard}} \\
& & & CER $\downarrow$ & SIM $\uparrow$
& WER $\downarrow$ & SIM $\uparrow$
& CER $\downarrow$ & SIM $\uparrow$ \\
\midrule
None & -- & -- & 2.17 & 0.697 & 4.63 & 0.600 & 20.13 & 0.656 \\
Token Emb. & \(\mathcal{L}_{\cos}\) & 0.1 & 2.12 & \underline{0.709} & 4.62 & 0.614 & 19.43 & 0.665 \\
Token Emb. & \(\mathcal{L}_{\cos}\) & 0.5 & 2.01 & \underline{0.709} & 4.13 & 0.618 & 19.05 & \underline{0.667} \\
Token Emb. & \(\mathcal{L}_{\cos}\) & 1.0 & \underline{1.90} & \textbf{0.711} & \underline{3.68} & \underline{0.622} & \underline{18.86} & \textbf{0.669} \\
Token ID & \(\mathcal{L}_{\mathrm{CE}}\) & 0.1 & \textbf{1.47} & \textbf{0.711} & \textbf{2.81} & \textbf{0.631} & \textbf{15.94} & 0.664 \\
\bottomrule
\end{tabular}%
}
}
\caption{
Comparison between continuous token-embedding regression with cosine loss and
discrete token-ID classification with cross-entropy loss. The unanchored model
serves as the baseline. Best and second-best results are shown in
\textbf{bold} and \underline{underlined}.
}
\label{tab:supp_anchoring_objective}
\end{table}

\paragraph{Embedding Regression versus Token Classification.}
Table~\ref{tab:supp_anchoring_objective} compares continuous token-embedding
regression based on cosine loss with discrete token-ID classification based on
cross-entropy loss. Both objectives use targets from the same frozen semantic
tokenizer and are applied to the same Transformer block. All other model,
data, optimization, and inference settings are identical.

For continuous embedding regression, increasing
\(\lambda_{\mathrm{sem}}\) from \(0.1\) to \(1.0\) consistently reduces all
three content-error metrics without decreasing SIM. Therefore,
\(\lambda_{\mathrm{sem}}=1.0\) is the best continuous-target configuration
among those evaluated. Nevertheless, discrete token-ID classification with
\(\lambda_{\mathrm{sem}}=0.1\) further reduces ZH CER, EN WER, and ZH-Hard
CER from \(1.90\), \(3.68\), and \(18.86\) to \(1.47\), \(2.81\), and
\(15.94\), corresponding to relative reductions of \(22.6\%\), \(23.6\%\),
and \(15.5\%\), respectively. Compared with the unanchored baseline, the
relative reductions are \(32.3\%\), \(39.3\%\), and \(20.8\%\),
respectively. Token-ID classification also matches the best continuous
variant on ZH SIM and improves EN SIM from \(0.622\) to \(0.631\), although
ZH-Hard SIM decreases slightly from \(0.669\) to \(0.664\). Under the matched
setting, these results indicate that discrete token-ID classification provides
a more effective anchoring objective for content modeling than continuous
embedding regression.

\begin{figure}[t]
\centering
\includegraphics[
    width=0.85\linewidth,
    height=0.24\textheight,
    keepaspectratio
]{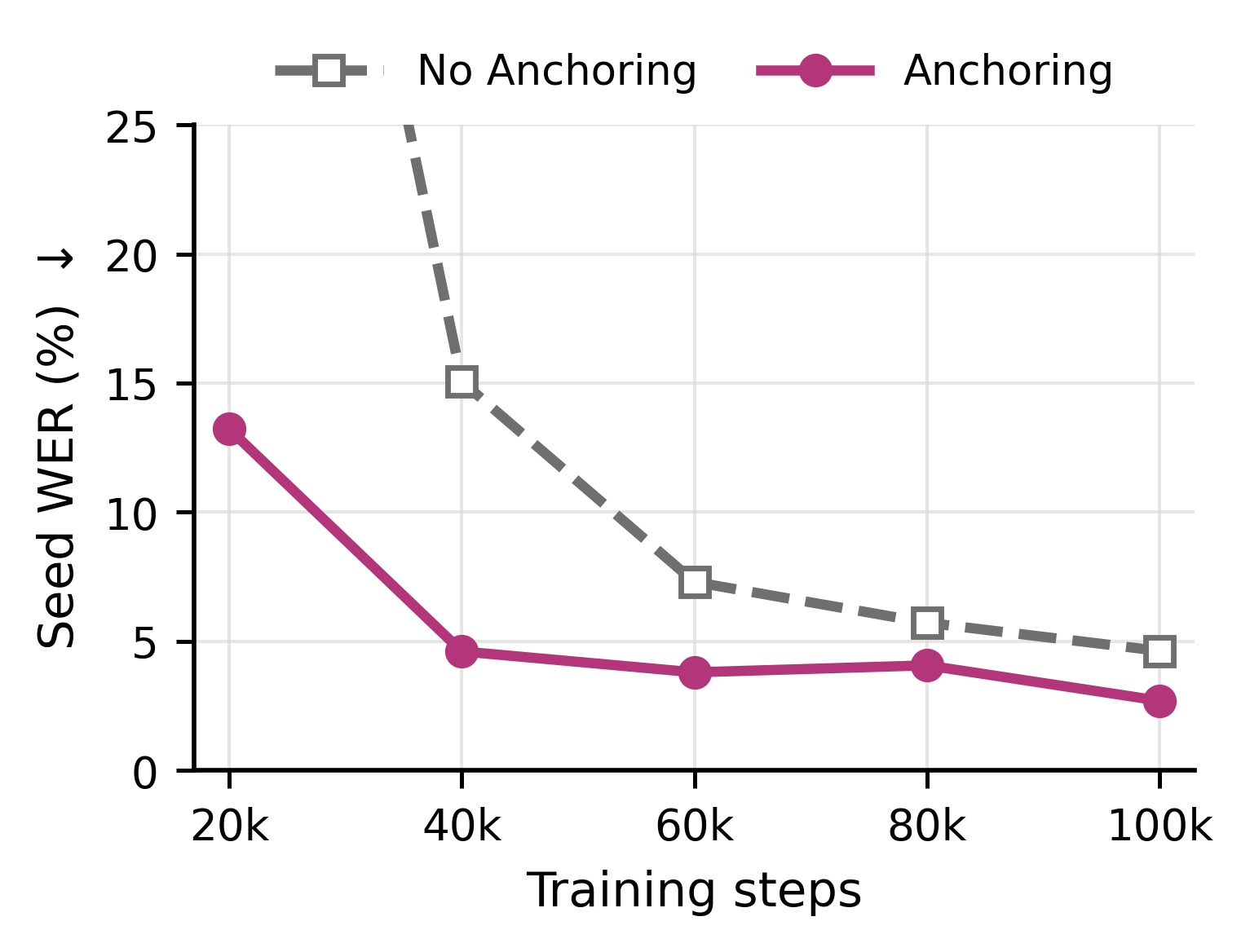}
\caption{
Seed-TTS EN WER across training updates with and without
semantic-token anchoring.
}
\label{fig:supp_anchoring_wer_dynamics}
\end{figure}

\begin{figure}[t]
\centering
\includegraphics[
    width=0.85\linewidth,
    height=0.24\textheight,
    keepaspectratio
]{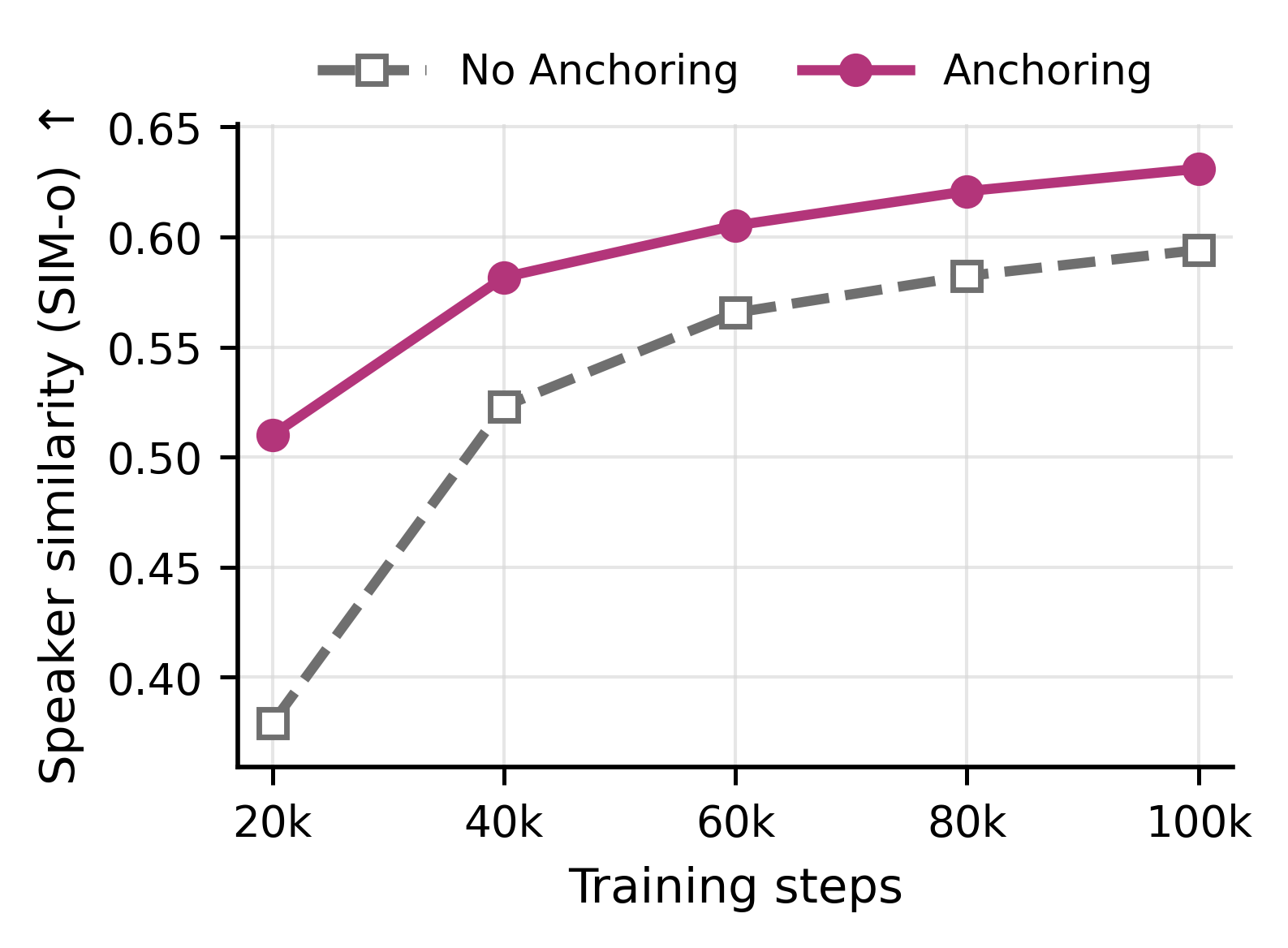}
\caption{
Seed-TTS EN SIM-o across training updates with and without
semantic-token anchoring.
}
\label{fig:supp_anchoring_sim_dynamics}
\end{figure}

\paragraph{Convergence Behavior with Semantic-Token Anchoring.}
We evaluate both configurations every 20K updates using the Seed-TTS EN
streaming protocol with a CFG scale of 2.0. At each checkpoint, we perform one
inference run on all 1,088 utterances. As shown in
Figures~\ref{fig:supp_anchoring_wer_dynamics}
and~\ref{fig:supp_anchoring_sim_dynamics}, semantic-token anchoring achieves
lower WER and higher speaker similarity at every evaluated checkpoint.

By 40K updates, the anchored model achieves a WER of \(4.61\%\) and a SIM-o
score of \(0.5817\), approaching the performance of the unanchored model after
100K updates: its WER is already slightly lower (\(4.61\%\) vs.\ \(4.63\%\)),
while its SIM-o remains close (\(0.5817\) vs.\ \(0.5941\)). 
At 100K updates, semantic-token anchoring reduces WER from \(4.63\%\) to
\(2.70\%\), corresponding to a relative reduction of \(41.7\%\), and
increases SIM-o from \(0.5941\) to \(0.6310\). The WER curve is not strictly
monotonic because of a small fluctuation between 60K and 80K updates, whereas
SIM-o improves monotonically throughout training. These results indicate that
semantic-token anchoring accelerates convergence and improves the balance
between intelligibility and speaker similarity. For clarity, the WER axis is
limited to \(25\%\). The WER of the unanchored model at 20K updates is
\(58.38\%\).


\subsection{Linguistic Probing and Representation Visualization.}

\begin{figure}[t]
\centering
\includegraphics[width=\linewidth]{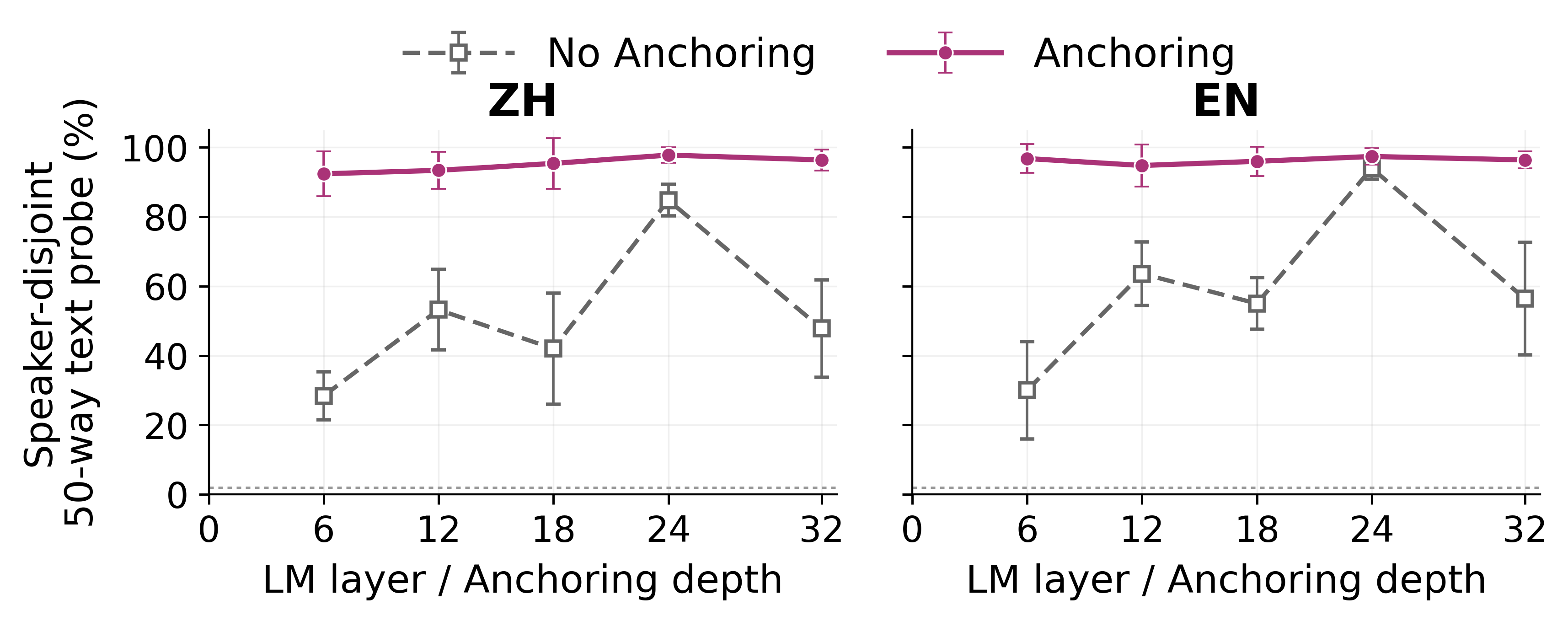}
\caption{Speaker-disjoint target-text probing across LM layers and Anchoring depths.}
\label{fig:supp_text_probe}
\end{figure}

\begin{figure}[t]
\centering
\includegraphics[width=\columnwidth]{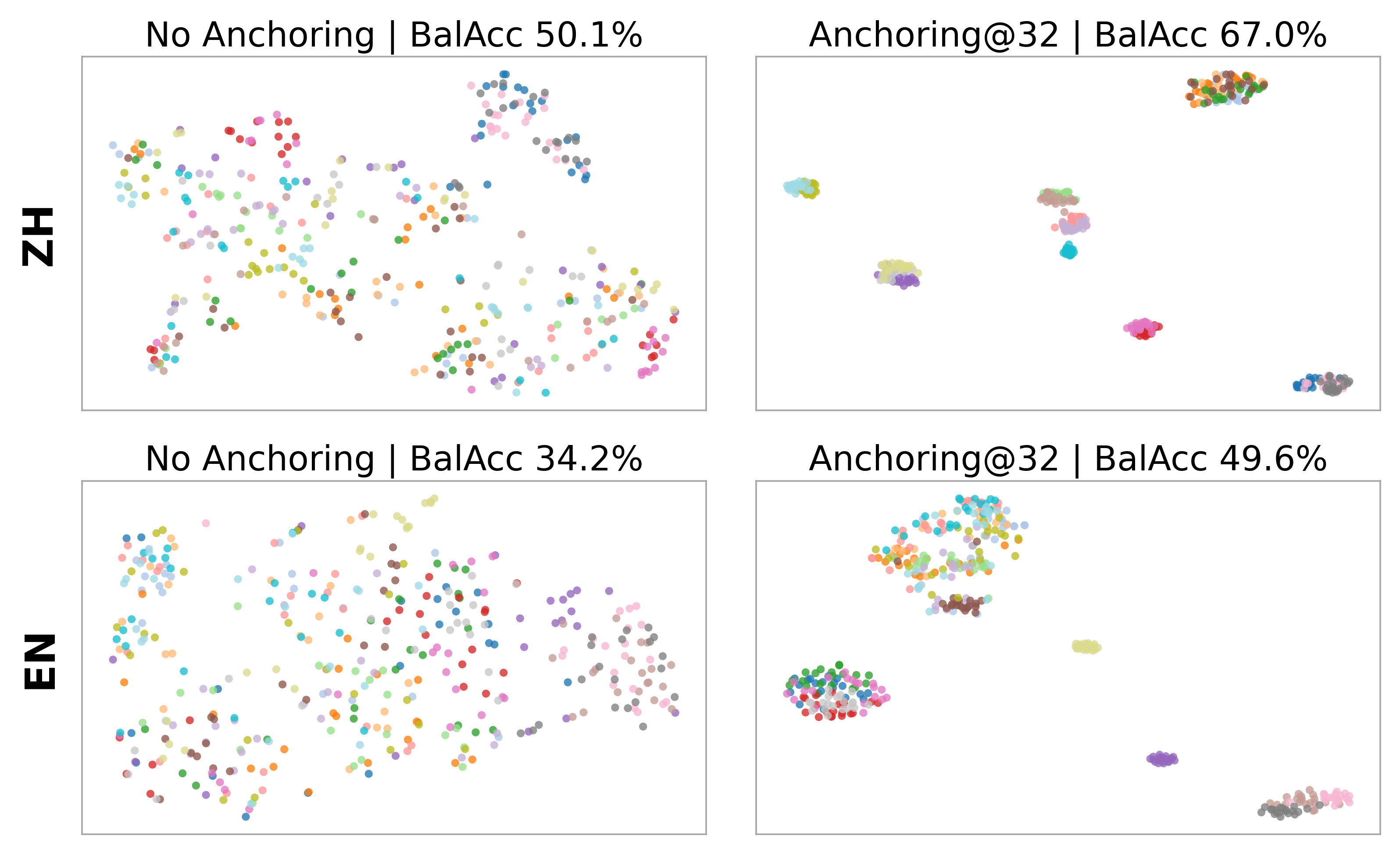}
\caption{Semantic-token t-SNE at \(h_{32}\) for No Anchoring and Anchoring@32.}
\label{fig:supp_noce_ce32_tsne}
\end{figure}

\begin{figure}[t]
\centering
\includegraphics[width=\columnwidth]{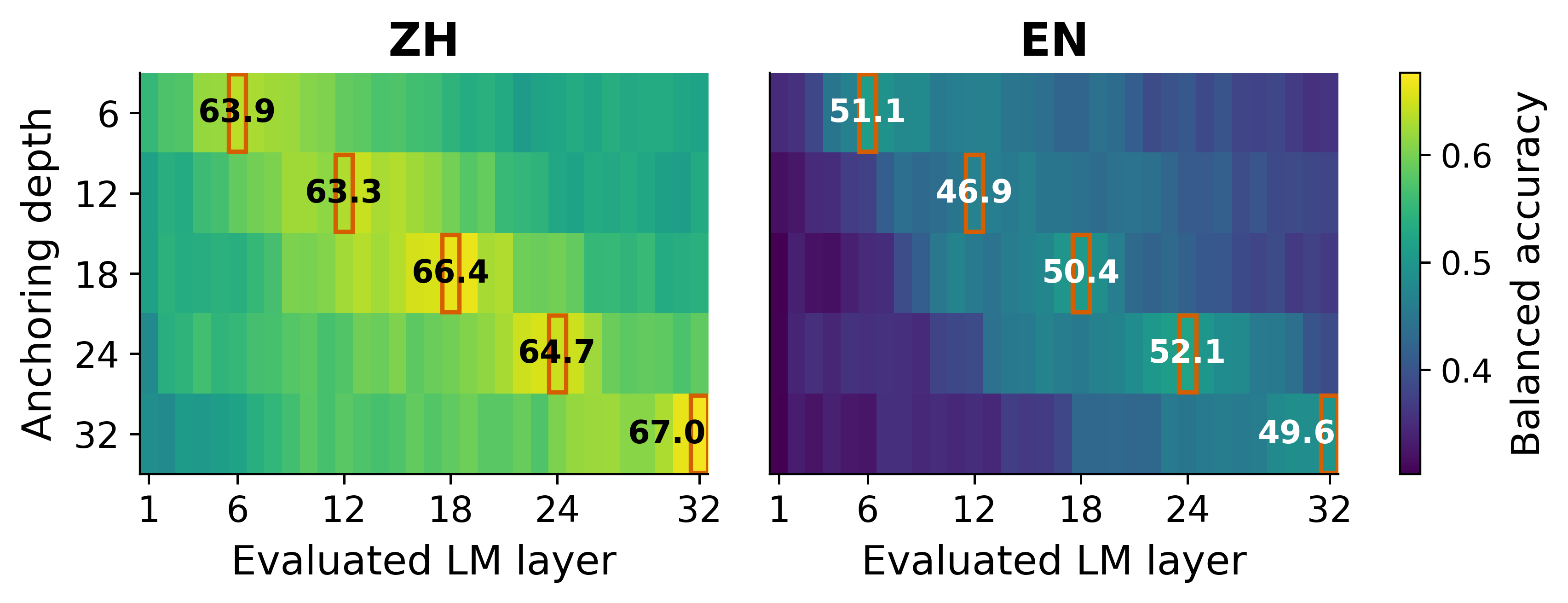}
\caption{Layer-wise semantic-token readability under different Anchoring depths.}
\label{fig:supp_anchoring}
\end{figure}

\paragraph{Speaker-Disjoint Linguistic Probing.}
We investigate whether target linguistic content remains linearly accessible
independently of prompt-speaker identity with a speaker-disjoint 50-way
text-identification probe. For each language, the probe uses mean-pooled hidden
states over target-generation positions from 10 prompt speakers and 50 shared
target texts. Prompt speakers are partitioned by five-fold GroupKFold so no
speaker appears in both training and test splits.

As shown in Figure~\ref{fig:supp_text_probe}, models anchored at tested
depths achieve consistently high text-identification accuracy, ranging from
\(92.4\%\) to \(97.8\%\). In contrast, the unanchored baseline exhibits a
strongly non-monotonic layer-wise profile. For Mandarin and English,
respectively, accuracy begins at \(28.4\%\) and \(30.0\%\) at \(h_6\),
increases to \(84.8\%\) and \(94.0\%\) at \(h_{24}\), then decreases to
\(47.8\%\) and \(56.4\%\) at \(h_{32}\). The maximum at \(h_{24}\) coincides
with the boundary between the 24-layer base LM and the subsequent eight-layer
residual LM. This pattern suggests that linguistic content is naturally
accessible at this interface but becomes less linearly separable after the
residual computation. Semantic-token anchoring does not introduce linguistic
information absent from the baseline. Instead, it makes this information
accessible at earlier depths and preserves high readability at the supervised
layer. The anchoring curve connects independently trained models with different
supervision depths and should be interpreted as a depth-ablation trend rather
than a layer-wise trajectory within a single model.

\begin{table*}[t]
\centering
\footnotesize
\setlength{\tabcolsep}{2.2pt}
\renewcommand{\arraystretch}{0.93}

\begin{tabular}{@{}lc|cccc|cccc@{}}
\toprule
\multirow{2}{*}{\textbf{System}} &
\multirow{2}{*}{\textbf{Control}} &
\multicolumn{4}{c|}{\textbf{Mandarin}} &
\multicolumn{4}{c}{\textbf{English}} \\
\cmidrule(lr){3-6}\cmidrule(lr){7-10}
& & CER (\%) $\downarrow$ & SIM $\uparrow$ & SingMOS $\uparrow$ & Sheet $\uparrow$ & WER (\%) $\downarrow$ & SIM $\uparrow$ & SingMOS $\uparrow$ & Sheet $\uparrow$ \\
\midrule
\multicolumn{10}{l}{\textit{Existing SVS systems}} \\
StyleSinger & Score & 38.80 & 0.808 & 3.892 & 3.833 & -- & -- & -- & -- \\
TCSinger & Score & 22.70 & 0.822 & 3.939 & 4.009 & 46.00 & 0.729 & 3.652 & 3.685 \\
Ying-Music-Singer & Melody & 11.70 & 0.913 & 4.068 & 3.599 & -- & -- & -- & -- \\
Vevo & Melody & 23.30 & 0.908 & 4.265 & 3.814 & 25.60 & 0.888 & 4.184 & 3.582 \\
SoulX-Singer & Score & \textbf{6.90} & 0.922 & 4.394 & 4.053 & 12.90 & \underline{0.914} & 4.255 & 3.690 \\
\midrule
\multicolumn{10}{l}{\textit{Continuous-latent autoregressive system}} \\
\textbf{\methodname{}} & Score & 10.71 & \textbf{0.925} & \textbf{4.470} & \textbf{4.224} & \textbf{9.50} & \underline{0.914} & \textbf{4.456} & \textbf{4.088} \\
\midrule
w/o Anchoring & Score & 11.56 & \underline{0.922} & \underline{4.455} & \underline{4.201} & \underline{11.42} & \textbf{0.916} & \underline{4.453} & \underline{3.981} \\
\bottomrule
\end{tabular}
\caption{Results on SoulX-Singer-Eval. Control indicates whether each system
uses melody or score conditioning. Lower CER/WER is better; higher SIM, SingMOS, and Sheet-SSQA are
better.}
\label{tab:supp_soulx_singer}
\end{table*}

\paragraph{Semantic-Token Geometry at the Final Layer.}
We examine the patch-level organization of semantic information in the final
LM layer. For each language, we select 20 frequent next-semantic-token classes
and sample 20 valid acoustic-patch states for each class. We use the
training-time causal alignment: the hidden state at acoustic patch \(k\) is
paired with semantic token \(k+1\).

Figure~\ref{fig:supp_noce_ce32_tsne} visualizes the \(h_{32}\)
representations after \(\ell_2\) normalization, PCA reduction, and a joint
t-SNE projection fitted to both configurations. With Anchor@32, balanced
accuracy for semantic-token prediction increases from \(50.1\%\) to
\(67.0\%\) for Mandarin and from \(34.2\%\) to \(49.6\%\) for English. The
anchored representations form more compact, class-specific clusters, whereas
the unanchored representations remain broadly intermixed. Overlap remains
because each LM state jointly encodes linguistic, speaker, and acoustic
context, and nearby semantic tokens may represent phonetically related
content. We treat the t-SNE visualization as qualitative evidence of a change
in local representation geometry. The speaker-disjoint probe in the original
representation space provides the primary quantitative evidence.

\paragraph{Depth-Localized Semantic Readability.}
To examine how supervision depth reshapes the LM hierarchy, we probe normalized
hidden states from every LM layer in independently trained Anchor@k models.
Figure~\ref{fig:supp_anchoring} reveals a depth-localized pattern: the region
of highest semantic-token readability shifts toward the supervised layer
rather than increasing uniformly with network depth. At the supervised layers,
balanced accuracy ranges from \(63.3\%\) to \(67.0\%\) for Mandarin and from
\(46.9\%\) to \(52.1\%\) for English.

This layer-following pattern indicates that the auxiliary objective shifts
where next-token semantic information is most linearly accessible within the
causal LM. However, accuracy at the supervised layer is not monotonic with
anchoring depth, and the deepest configuration is not consistently strongest
across languages. These results suggest that semantic-token anchoring
localizes semantic accessibility around the selected supervision depth, rather
than showing that progressively deeper anchoring is always preferable for
generation.

\subsection{Inference-Time Classifier-Free Guidance}
\label{sec:supp_cfg_analysis}

\begin{figure}[t]
\centering
\includegraphics[width=\columnwidth]
{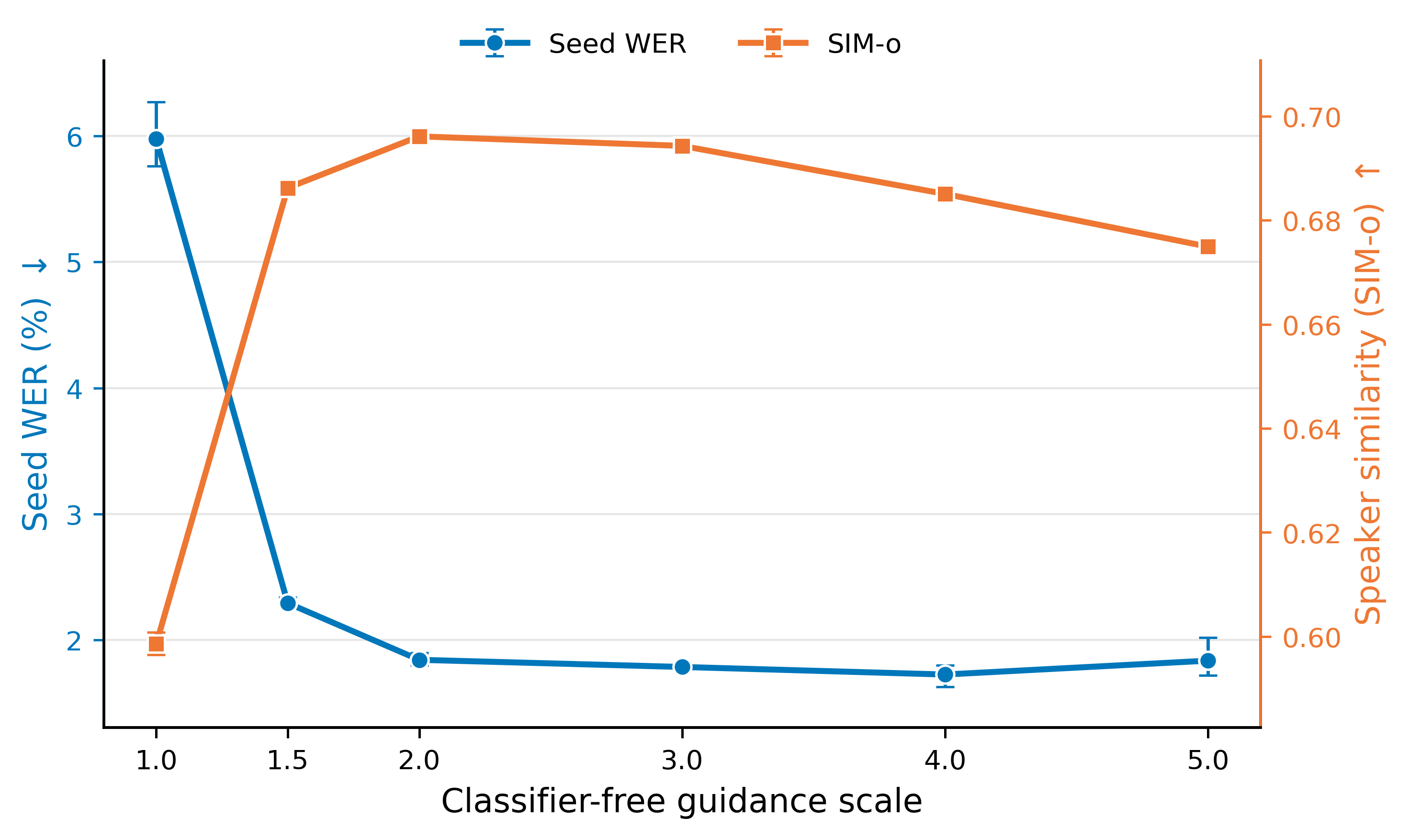}
\caption{CFG-scale trade-off on streaming Seed-TTS EN at 300k steps, averaged over three runs.}
\label{fig:supp_cfg_tradeoff}
\end{figure}

Figure~\ref{fig:supp_cfg_tradeoff} evaluates the effect of classifier-free
guidance (CFG) on streaming Seed-TTS EN generation. Increasing the guidance
scale from \(1.0\) to \(2.0\) substantially reduces Seed WER from \(5.98\%\)
to \(1.84\%\), while improving speaker similarity from \(0.599\) to \(0.695\).
Stronger guidance between \(3.0\) and \(4.0\) provides only modest additional
improvements in intelligibility, accompanied by a consistent reduction in
speaker similarity. At \(5.0\), WER no longer improves and similarity further
decreases. These results reveal a clear intelligibility--speaker-similarity
trade-off, with CFG \(=2.0\) providing the most balanced operating point.

\subsection{Extension to Score-Conditioned SVS}
\label{sec:supp_svs_results}

\paragraph{Results on SoulX-Singer-Eval.}
\label{sec:supp_soulx_results}

Together with the GMO-SVS results, SoulX-Singer-Eval demonstrates that the
continuous-latent formulation remains competitive under symbolic musical
control. Under the matched SoulX-Singer-Eval setting, semantic-token anchoring
reduces Mandarin CER from 11.56 to 10.71 and English WER from 11.42 to 9.50.
Singer similarity remains nearly unchanged, while SingMOS and Sheet are
maintained or improved in both languages. These results further support the
transfer of semantic-token anchoring to lyric modeling without materially
compromising singer identity or perceptual quality.

\section{Limitations}
\label{sec:supp_qualitative_limitations}




\methodname{} currently relies on semantic targets from a single frozen
tokenizer, and its effectiveness may therefore depend on the tokenizer's
temporal granularity, vocabulary, and domain coverage, especially for speech
styles or languages underrepresented during tokenizer pretraining. Each
configuration is trained once using fixed hyperparameters and a predefined
training seed. For evaluation, we perform three inference runs with the same
decoding settings and predefined sampling seeds, and report the averaged
results. This procedure reduces inference-time sampling variation but does not
characterize variation across independently trained models. Moreover,
automatic metrics do not fully replace human evaluation, thereby limiting
direct claims about perceptual quality and listener preference; the scaled
120K-hour setting changes both data scale and task composition, and comparisons
with externally reported systems are not fully matched. Our conclusions should
therefore be interpreted within the evaluated settings.

As with other high-fidelity voice-generation systems, \methodname{} may be
misused for impersonation or deceptive content. Its deployment should comply
with applicable data licenses, speaker consent, privacy requirements, and
responsible-use practices, including appropriate access control and disclosure
of generated content. The internal singing data cannot be fully redistributed
because of licensing restrictions.

\end{document}